\documentclass[12pt]{article}
\usepackage{amsmath}
\usepackage{graphicx}
\usepackage{enumerate}
\usepackage{natbib}
\usepackage{url} 
\usepackage{rotating}
\newcommand{\blind}{1}

\usepackage[utf8]{inputenc}
\usepackage{bm}
\usepackage[hypertexnames=false]{hyperref}
\usepackage{algorithm,algpseudocode}
\usepackage{amsthm,amssymb,amsfonts}
\usepackage{thmtools}
\usepackage{chngcntr}
\usepackage{mathrsfs}
\usepackage{multirow}
\usepackage{verbatim}
\usepackage{titlesec,titling}
\usepackage{bbm}
\usepackage{subcaption}
\usepackage{enumitem}
\newtheorem{theorem}{Theorem}
\newtheorem{lemma}{Lemma}
\newtheorem{corollary}{Corollary}
\newtheorem{assumption}{Assumption}
\newtheorem{definition}{Definition}
\declaretheoremstyle[notefont=\normalfont]{normalhead}
\declaretheorem[style=normalhead,qed=$\square$]{example}
\theoremstyle{remark}

\usepackage[super]{nth}
\def\ind{\perp\!\!\!\!\perp}
\renewcommand{\baselinestretch}{1.3}
\usepackage{xcolor}

\begin{document}

\def\spacingset#1{\renewcommand{\baselinestretch}%
{#1}\small\normalsize} \spacingset{1}


\if1\blind
{
  \title{\bf An additive graphical model for discrete data}
  \author{Jun Tao, Bing Li, and Lingzhou Xue\\
Department of Statistics, Pennsylvania State University}
  \date{First Version: October, 2020\\ This Version: December, 2021}
  \maketitle
} \fi

\if0\blind
{
  \bigskip
  \bigskip
  \bigskip
  \begin{center}
    {\LARGE\bf An additive graphical model for discrete data}
\end{center}
  \medskip
} \fi

\bigskip
\begin{abstract}
We introduce a nonparametric graphical model for discrete node variables based on additive conditional independence. Additive conditional independence is a three way statistical relation that shares similar properties with conditional independence by satisfying the semi-graphoid axioms. Based on this relation we build an additive graphical model for discrete variables that does not suffer from the restriction of a parametric model such as the Ising model. We develop an estimator of the new graphical model via the penalized estimation of the discrete version of the additive precision operator and establish the consistency of the estimator under the ultrahigh-dimensional setting. Along with these methodological developments, we also exploit the properties of discrete random variables to uncover a deeper relation between additive conditional independence and conditional independence than previously known. The new graphical model reduces to a conditional independence graphical model under certain sparsity conditions. We conduct simulation experiments and analysis of an HIV antiretroviral therapy  data set to compare the new method with existing ones.
\end{abstract}

\noindent%
\vfill

\section{Introduction}
\label{sec:intro}
A graphical model uses a graph-based representation as the basis for analyzing multivariate data.
It has been gaining popularity in various applied fields. The most frequently studied graphical models are those based on conditional independence (CI) between node variables, which are also called Markov random fields (MRF). Let
$X = (X^1,X^2,\cdots,X^p)^\top$ be a $p$-dimensional random vector, $\mathsf{V}=\{1,\cdots,p\}$, and $\mathcal{E} \subseteq\{(i,j)\in\mathsf{V}\times\mathsf{V} :i\ne j\}$. Then $X$ follows a Markov random field with respect to the undirected graph $\mathcal{G} = \{\mathsf{V}, \mathcal{E}\}$ if and only if
$X^i$ and $X^j$ are independent given the rest of $X$, that is,
\begin{equation}\label{cigraph}
    (i,j)\notin \mathcal{E} \Leftrightarrow X^i\ind X^j \mid X^{-\{i,j\}} ,
\end{equation}
where $X^{-\{i,j\}}$ represents $X$ with its $i$-th and $j$-th components removed.

 Discrete graphical models, where $X^1,X^2,\cdots,X^p$ are discrete random variables, naturally appear in many fields. The simplest of such models is the binary graphical model, where the node vector $X$ is a member of $\{a_0,a_1\}^p$, with $a_0$ and $a_1$ being the possible labels. One particular example is the Ising model, where $X \in \{-1,1\}^p$.
The Ising model (\cite{ising1925beitrag}) was originally introduced as a mathematical model of ferromagnetism in statistical mechanics. Nowadays it has been applied to such diverse fields as image restoration (\cite{geman1993stochastic}), biophysics (\cite{ahsan1998elasticity}), genetics (\cite{fierst2015modeling}), and network psychometrics (\cite{marsman2018introduction}).

The Ising model has the probability mass function (p.m.f.) of the form:
\begin{equation}\label{ising_def}
    f_{\bm{\beta}}(x^1,x^2,\cdots,x^p) =\frac{1}{z(\bm{\beta})} \exp\Big(\sum_{i}\beta_{ii}x^i+\sum_{i<j}\beta_{ij}x^ix^j\Big),\end{equation}
where $z(\bm{\beta})= \sum_{x\in \{-1,1\}^p}\exp\Big(\sum_{i}\beta_{ii}x^i+\sum_{i<j}\beta_{ij}x^ix^j\Big)$ is the partition function.
Let $\beta_{ij} := \beta_{ji}$, for $i>j$. It is easier to view the parameter as the symmetric matrix $\bm{\beta} = (\beta_{ij})_{i,j=1}^p$ in $\mathbb{R}^{p\times p}$. For any pair $(i,j)$ with $i\ne j$, the parameter $\beta_{ij}$ characterizes the conditional independence between
$X^i$ and $X^j$ given $X^{-\{i,j\}}$ by the equivalence
\begin{equation}\label{isinggraph}
 X^i\ind X^j \mid  X^{-\{i,j\}}\Leftrightarrow \beta_{ij}=0,
\end{equation}
as can be seen from the conditional p.m.f.
\[f(x^i,x^j\mid x^{-\{i,j\}}) \propto \exp\Big(\beta_{ij}x^ix^j+x^i(\beta_{ii}+\sum_{k\ne i,j}\beta_{ik}x^k)+x^j(\beta_{jj}+\sum_{k\ne i,j}\beta_{jk}x^k)\Big).\]
Wherever $\beta_{ij}=0$, the conditional p.m.f. is separable for $x^i$ and $x^j$.
Thus, by \eqref{cigraph} and \eqref{isinggraph}, for the Ising model, estimating $\mathcal{G}$ amounts to identifying the zero entries or, equivalently, the sparsity pattern of the symmetric matrix $\bm{\beta}$. There have been some related works for learning the edge set of the Ising model. For example, \cite{hofling2009estimation}, \cite{wang2011learning} and \cite{xue2012nonconcave} developed the penalized estimation procedure based on a pseudo-likelihood, while \cite{ravikumar2010high} suggested fitting separate $\ell_1$-penalized logistic regressions for each node to learn its neighborhood. \cite{cheng2014sparse} proposed a sparse covariate dependent Ising model. \cite{guo2015graphical} and \cite{lee2021estimating} studied the graphical models for discrete and ordered data.


For many applications, the Ising distribution assumption can be violated, making the Ising model inapplicable. An intuitive explanation is that an Ising model has limited degrees of freedom: it can only explain $p(p+1)/2$ out of a total of $(2^p-1)$ degrees of freedom in the probability mass function. Thus, for a large graph, the Ising models often become inadequate. In this paper, we seek for a new discrete graphical model without the Ising distribution assumption, which retains the fundamental simplicity of the Ising dependence structure in \eqref{isinggraph}. This is achieved by replacing conditional independence by additive conditional independence (ACI), which is a new statistical relation introduced by \cite{li2014additive}. Inspired by this idea, we propose a discrete graphical model derived from ACI, which we refer to as the discrete additive semi-graphoid model (DASG).

Our methods can be summarized as follows.
Between the Hilbert spaces of the functions of the node variables, we introduce the cross-covariance operators and use them to define a discrete additive precision operator (DAPO). The DAPO is a matrix of linear operators that inherits the relation \eqref{isinggraph} at the operator level: the $(i, j)$-th entry of the DAPO is the zero operator if and only if $X^i$ and $X^j$ are additively conditionally independent given $X^{-\{i,j\}}$. The nonzero entries of the DAPO then determine the edges of the graph.

\cite{li2014additive} and \cite{lee2016additive} developed an estimator based on the additive precision operator (APO) to learn the edge set of the ACI graphs with continuous valued node variables. The APO-based estimator is a hard-thresholding estimator, which asymptotically converges at a relatively slow rate. Furthermore, the tuning of the thresholding parameter is based on generalized cross validation, which can be difficult to carry out thoroughly when $p$ is very large.
There are some important differences between discrete graphs and continuous ones. For example, in the discrete case, the Hilbert spaces of node functions are finite dimensional and the properties of the related operators—particularly those pertaining to their inverses—are a lot more pleasant. We use a penalized method to obtain a sparse matrix estimation of the DAPO, which has faster convergence rate. Additionally, our estimator can be easily tuned with common tuning methods like cross validations.
It is worth pointing out that we also investigate the relation between ACI and CI to a deeper degree than previously understood, which fills a gap in the literature. This result is made possible by carefully studying the special structure of the discrete distribution.

The rest of the article is organized as follows. In Section \ref{s2}, we introduce the DASG and its operator characterization. In Section \ref{s3}, we investigate the relation between ACI and CI under the Ising model. In Section \ref{s4}, we derive sample-level coordinate representations of DASG linear operators and develop the new estimator. In Section \ref{s5}, we establish the asymptotic properties of the proposed estimator. In Section \ref{s6}, we conduct simulation experiments to evaluate our estimator. In Section \ref{s7}, we apply our estimator to an HIV antiretroviral therapy dataset. All the proofs are presented in the supplementary material.

\section{Discrete additive semi-graphoid model}\label{s2}
We first review the definition of additive conditional independence and  propose the discrete additive semi-graphoid model in Subsection 2.1. We then introduce the equivalent form of additive conditional independence in terms of linear operators in Subsection 2.2.
\subsection{Additive conditional independence}

Let $\mathscr{R}\subseteq 2^\mathsf{V}\times2^\mathsf{V}\times2^\mathsf{V}$ be a three-way relation on $\mathsf{V}$, where $2^\mathsf{V}$ consists of all
subsets of $\mathsf{V}$.

\begin{definition} (\cite{pearl1987logic})
A three-way relation $\mathscr{R}$ is called a semi-graphoid if it satisfies the following conditions:
\begin{itemize}
    \item (symmetry) $(A, C, B) \in \mathscr{R}\Rightarrow (B,C,A) \in \mathscr{R}$;
    \item (decomposition) $(A, C, B \cup D) \in \mathscr{R}\Rightarrow (A, C, B) \in \mathscr{R}$;
    \item (weak union) $(A, C, B \cup D) \in \mathscr{R}\Rightarrow (A, C\cup B, D) \in \mathscr{R}$;
    \item (contraction) $(A, C\cup B, D) \in \mathscr{R},\ (A, C,B) \in \mathscr{R} \Rightarrow (A, C, B \cup D) \in \mathscr{R}$.
\end{itemize}
\end{definition}
These axioms are extracted from the conditional independence (CI) to convey the general idea of ``$B$ is irrelevant for understanding $A$ once $C$ is known," or ``$C$ separates $A$ and $B$." The  CI is a special case of the semi-graphoid three-way relation (\cite{dawid1979conditional}).

\cite{li2014additive} proposed the notion of the additive conditional independence (ACI) as a promising alternative to the CI for constructing graphical models. The ACI satisfies the axioms for a semi-graphoid (\cite{pearl1987logic}), which captures the essence of a graph.

Let $X = (X^1,X^2,\cdots,X^p)^\top\in\mathbb{R}^p$ be a random vector. For any node $i\in \mathsf{V}$, let $L^2(P_{X^i})$ denote the class of functions of $X^i$ such that $E\phi(X^i) = 0,\ E\phi^2(X^i) < \infty$. Let $\mathscr{A}_{X^i}\subseteq L^2(P_{X^i})$ be a Hilbert subspace. For any nonempty node subset $A\subseteq \mathsf{V}$, the subvector of $X$ on $A$ means $X^A := \left\{X^i\right\}_{i \in A}$. Let $\mathscr{A}_{X^A}$ denote the additive family
\[\sum_{i\in A}\mathscr{A}_{X^i}=\Big\{\sum_{i\in A}\phi_i: \phi_i\in \mathscr{A}_{X^i},\ i\in A\Big\}.\]
For two subspaces $\mathscr{A}$ and $\mathscr{B}$ of $L^2(P_X)$, let $\mathscr{A}\ominus\mathscr{B} = \mathscr{A}\cap\mathscr{B}^\perp$, where the orthogonality is in terms of the $L^2(P_X)$-inner product.

\begin{definition}\label{aci_def}
Let $X^A, X^B,$ and $X^C$ be subvectors of $X$. $X^A$ and $X^B$ are additively conditionally independent  given $X^C$ with respect to $(\mathscr{A}_{X^A},\mathscr{A}_{X^B},\mathscr{A}_{X^C})$ if and only if
\[(\mathscr{A}_{X^A}+\mathscr{A}_{X^C})\ominus\mathscr{A}_{X^C}\perp (\mathscr{A}_{X^B}+\mathscr{A}_{X^C})\ominus\mathscr{A}_{X^C},\]
where $\perp$ indicates orthogonality with respect to the $L^2(P_X)$-inner product. The above relation is written as $X^A\ind_{A}X^B\mid X^C$.
\end{definition}

Suppose
$X = (X^1,X^2,\cdots,X^p)^\top\in\{0,1,\cdots,m\}^p$, $m \geqslant 1$.
The discrete additive semi-graphoid model (DASG) is defined through the following equivalence
$$X^i\ind_A X^j \mid  X^{-\{i,j\}}\Leftrightarrow (i,j)\notin \mathcal{E}.$$

\subsection{Discrete additive precision operator}
We now introduce a linear operator that characterizes the ACI.

\begin{definition}
(\cite{baker1973joint, fukumizu2009kernel})
The cross-covariance operator of $(X^j,X^i)$, $\Sigma_{X^iX^j}$, is a mapping from $\mathscr{A}_{X^j}$ to $\mathscr{A}_{X^i}$, such that for any $\phi\in \mathscr{A}_{X^i},\ \psi\in \mathscr{A}_{X^j}$
\begin{equation}\label{cross-cov}
    \langle\phi, \Sigma_{X^iX^j}\psi\rangle_{\mathscr{A}_{X^i}} = \mathrm{cov}[\phi(X^i),\psi(X^j)].
\end{equation}
\end{definition}

By the Cauchy-Schwarz inequality and $X$ being finitely discrete, the bilinear form $\mathscr{A}_{X^i}\times\mathscr{A}_{X^j} \rightarrow \mathbb{R}$: $(\phi,\psi) \mapsto \mathrm{cov}[\phi(X^i),\psi(X^j)]$ is bounded. The existence and uniqueness of $\Sigma_{X^iX^j}$ defined through \eqref{cross-cov} is guaranteed by Riesz’s representation theorem.

Let $\mathrm{ker}(\Sigma_{X^i X^i})=\big\{h\in\mathscr{A}_{X^i}:\Sigma_{X^i X^i}h=0\big\}$ be the kernel space of $\Sigma_{X^i X^i}$. Note that $\phi \in \mathrm{ker}( \Sigma_{X^i X^i} )$ if and only if $\mathrm{var}[ \phi (X^i)]=0$. This implies that $\mathrm{ker}(\Sigma_{X^i X^i})$ is a linear subspace of constant functions. Since $\mathscr{A}_{X^i}$ only contains mean zero functions, we have $\mathrm{ker}(\Sigma_{X^i X^i} )=\{0\}$, which implies that the operator $\Sigma_{X^i X^i}$ is invertible.

The following definitions contain two key matrices of operators in our theory. 
\begin{definition}
 The operator $\Sigma_{XX}:=\big\{\Sigma_{X^iX^j}\big\}_{i,j=1}^p$ that satisfies $\Sigma_{XX}\phi =\sum_{i,j=1}^p \Sigma_{X^iX^j}\phi_j $ for any $\phi = \phi_1+\cdots+\phi_p\in \mathscr{A}_X$ is called the discrete additive variance operator (DAVO).

\end{definition}
By this definition, for any $\phi = \phi_1+\cdots+\phi_p$, $\psi = \psi_1+\cdots+\psi_p\in \mathscr{A}_X$,

$$\langle\phi, \Sigma_{XX}\psi\rangle_{\mathscr{A}_X} =\langle\phi, \sum_{i,j=1}^p \Sigma_{X^iX^j}\psi_j\rangle_{\mathscr{A}_X} =  \sum_{i,j=1}^p\langle\phi_i, \Sigma_{X^iX^j}\psi_j\rangle_{\mathscr{A}_{X^i}} $$
By the definition of cross-covariance operator, this is nothing but $\sum_{i,j=1}^p\mathrm{cov}[\phi_i(X^i),\psi_j(X^j)]$, which sums up to $\mathrm{cov}[\phi(X),\psi(X)]$. The variance matrix, or the covariance matrix, can be regarded as the Gram matrix of the set of random variables $\{X^1,\cdots,X^p\}$ with respect to the $L^2(P_X)$-inner product. Similarly, the DAVO serves as the Gram matrix of operators with respect to the inner product on the additive function space $\mathscr{A}_{X}$.

\begin{assumption}
$\mathrm{ker}(\Sigma_{XX})=\{0\}$.
\end{assumption}
This condition is satisfied by any non-degenerate node variables, where any nonzero function of the node variables will be linearly independent. Under this assumption, the mapping $\phi \mapsto \Sigma_{XX}\phi$ is invertible.

\begin{definition}
 The operator $\Sigma_{XX}^{-1}$ is called the discrete additive precision operator (DAPO), and is written as $\Theta_{XX}$.
\end{definition}

By being the inverse operator of $\Sigma_{XX}$, $\Theta_{XX}$ satisfies that $\phi=\Theta_{XX}\Sigma_{XX}\phi=\Sigma_{XX}\Theta_{XX}\phi$ for any $\phi\in\mathscr{A}_X$. We use $\Theta_{X^iX^j}$ to denote the $(i,j)$-th block of $\Theta_{XX}$, which is a mapping from $\mathscr{A}_{X^j}$ to $\mathscr{A}_{X^i}$, such that $\Theta_{XX}\phi =\sum_{i,j=1}^p \Theta_{X^iX^j}\phi_j$ for $\phi = \phi_1+\cdots+\phi_p$.
The next theorem re-expresses pairwise independence and ACI in terms of the DAVO and the DAPO.

\begin{theorem}\label{aci} The DAVO and the DAPO have the following properties
\begin{enumerate}[label=(\arabic*)]
    \item  $X^i\ind X^j$ if and only if $\Sigma_{X^iX^j}=0$,
    \item  $X^i\ind_A X^j \mid  X^{-\{i,j\}}$ if and only if $\Theta_{X^iX^j}=0$.
\end{enumerate}
\end{theorem}

The first statement of Theorem 1 gives a probabilistic interpretation of the DAVO, which can be shown by verifying the definition of independence. The relation between DAVO and the DASG is analogous to that between the covariance matrix and the Gaussian graphical model, where the covariance matrix determines pairwise independence, whereas its inverse determines CI. In our context, DAVO still determines pairwise independence, but its inverse, DAPO, determines the new statistical relationship, ACI.

\section{Relation between ACI and CI under the Ising model}\label{s3}
\cite{li2014additive} demonstrated
some relations between ACI and CI under the copula Gaussian
model assumption. \cite{loh2012structure} studied the relationship between the zero pattern of inverse covariance matrices and the edge set for Ising models. In this section, we further investigate the relation between ACI and CI under the Ising model, whose distributional simplicity allows us to uncover a deeper relation than previously understood.

\begin{definition} Consider a graph $\mathcal{G} = \{\mathsf{V}, \mathcal{E}\}$. For any node pair $(i,j)\in \mathsf{V}\times\mathsf{V}$ with $i\ne j$, a subset $R\subseteq \mathsf{V}\setminus\{i,j\}$ is an $(i,j)$-separator if the removal of the edge $(i,j)$ and the edges between $R$ and $\mathsf{V}\setminus R$ from the graph $\mathcal{G}$ separates $i$ and $j$ into distinct connected components.
\end{definition}

For any node pair $(i,j)$,
the set $\mathsf{V}\setminus\{i,j\}$ is naturally an $(i,j)$-separator, so the existence of a node separator is guaranteed. Given a node separator $R$, let $\mathcal{G}'$ be the graph with the edge $(i,j)$ and the edges between $R$ and $\mathsf{V}\setminus R$ removed. Consider the maximal connected subgraph of $\mathcal{G}'$ including $i$, whose node set is denoted by $C_i$. Similarly, let $C_j$ denote the node set of the maximal connected subgraph of $\mathcal{G}'$ including $j$. Then $C_i$ and $C_j$ are two mutually exclusive and disconnected components of the graph $\mathcal{G}'$.
\begin{definition} Let $X =(X^1,X^2,\cdots,X^p)^\top\in \mathbb{R}^p$ be a random node vector. For node $i\in\mathsf{V}$ and subset $D\subseteq \mathsf{V}\setminus\{i\}$, $\mathscr{A}_{X^i}$ is said to have linear conditional mean with respect to $\mathscr{A}_{X^D}$, if either $D=\varnothing$ or for any $\phi\in\mathscr{A}_{X^i}$,
\begin{equation}\label{lcm}
    E(\phi(X^i) \mid X^D) \in \mathscr{A}_{X^D}.
\end{equation}
\end{definition}
 When $D=\varnothing$, $X^D=\left\{X^j\right\}_{j \in D}$ is an empty subvector and $\mathscr{A}_{X^D}=\{0\}$. Then the meaning of linear conditional mean is that $E(\phi(X^i) \big)=0$ by the definition of $\mathscr{A}_{X^i}$. Let $|D|$ denote its cardinality.
When $X$ is binary, \eqref{lcm} means there exists a constant vector $\xi \in\mathbb{R}^{|D|}$ such that
\begin{equation*}
    E(X^i -EX^i\mid X^D) = \xi^\top (X^D-EX^D).
\end{equation*}
In this case, for simplicity, the binary node variable $X^i$ is said to have linear conditional mean with respect to $X^D$ directly, as $\mathscr{A}_{X^i}$ is one dimensional. The next theorem gives a relation between ACI and CI under a binary graphical model with linear conditional mean.

\begin{theorem}\label{separator}
Suppose $X = (X^1,X^2,\cdots,X^p)^\top\in \{a_0,a_1\}^p$, $p\geqslant 3$, follows a MRF with respect to $\{\mathsf{V}, \mathcal{E}\}$. For any node pair $(i,j)$ with $i\ne j$, if there exists an $(i,j)$-separator $R$ on $\{\mathsf{V}, \mathcal{E}\}$ such that $X^{k}$ has linear conditional mean with respect to $X^R$ for any $k\in C_i$, then
$X^i\ind X^j \mid  X^{-\{i,j\}}\Rightarrow X^i\ind_A X^j \mid  X^{-\{i,j\}}. $
\end{theorem}

Among the binary graphical models, the Ising model is of special interest. For simplicity, we focus on the family of symmetric Ising models, which is a special case of \eqref{ising_def} with $\beta_{ii}=0$, for $i=1,\cdots,p$. The corresponding p.m.f. is:
\begin{equation}\label{ising_sym}
f_{\bm{\beta}} (x^1,x^2,\cdots,x^p) =\frac{1}{z(\bm{\beta})} \exp\Big(\sum_{i<j}\beta_{ij}x^ix^j\Big).
\end{equation}
Under the symmetric Ising model, a more specific relation between ACI and CI can be obtained.
\begin{corollary}\label{isingsufficiency}
Suppose $X = (X^1,X^2,\cdots,X^p)^\top$, $p\geqslant 3$, follows a symmetric Ising model \eqref{ising_sym} with respect to $\{\mathsf{V}, \mathcal{E}\}$.
 For any node pair $(i,j)$ with $i\ne j$, if there exists an $(i,j)$-separator $R$ on  $\{\mathsf{V}, \mathcal{E}\}$ with $|R|  \leqslant 2$, then
$X^i\ind X^j \mid  X^{-\{i,j\}}\Rightarrow X^i\ind_A X^j \mid  X^{-\{i,j\}}. $
\end{corollary}

The equivalence between ACI and CI can be established if $R$ is chosen as the largest separator $\mathsf{V}\setminus\{i,j\}$ under the linear conditional mean assumption for this separator.

\begin{theorem}\label{isingequivalence}
Suppose $X = (X^1,X^2,\cdots,X^p)^\top$, $p\geqslant 3$, follows a symmetric Ising model \eqref{ising_sym} with respect to $\{\mathsf{V}, \mathcal{E}\}$. For any node pair $(i,j)$ with $i\ne j$, if $X^i$ has linear conditional mean with respect to $X^{-\{i,j\}}$, then
$X^i\ind X^j \mid  X^{-\{i,j\}}\Leftrightarrow X^i\ind_A X^j \mid  X^{-\{i,j\}}. $
\end{theorem}

However, the linear conditional mean can be easily violated by Ising models. To have a closer look at the reason, let $\mathcal{N}_i:=\big\{k\in\mathsf{V}\setminus\{i\}:(i,k)\in \mathcal{E}\big\}$ be the neighborhood of node $i\in\mathsf{V}$ on the given graph $(\mathsf{V}, \mathcal{E})$. Then $\mathcal{N}_i\setminus\{j\}$ and $\mathcal{N}_j\setminus\{i\}$ are $(i,j)$-separators as well, which can have a much smaller size than $\mathsf{V}\setminus\{i,j\}$.
\begin{corollary}\label{isingneighbor}
Suppose $X = (X^1,X^2,\cdots,X^p)^\top$, $p\geqslant 3$, follows a symmetric Ising model \eqref{ising_sym} with respect to $\{\mathsf{V}, \mathcal{E}\}$. For any node pair $(i,j)$ with $i\ne j$, if $\mathcal{N}_i$ is the neighborhood of node $i$ on $\{\mathsf{V}, \mathcal{E}\}$ and $|\mathcal{N}_i\setminus\{j\}|  \leqslant 2$, then
$X^i\ind X^j \mid  X^{-\{i,j\}}\Leftrightarrow X^i\ind_A X^j \mid  X^{-\{i,j\}}.  $
\end{corollary}

\begin{figure}[H]
\begin{center}
\includegraphics[width=15cm]{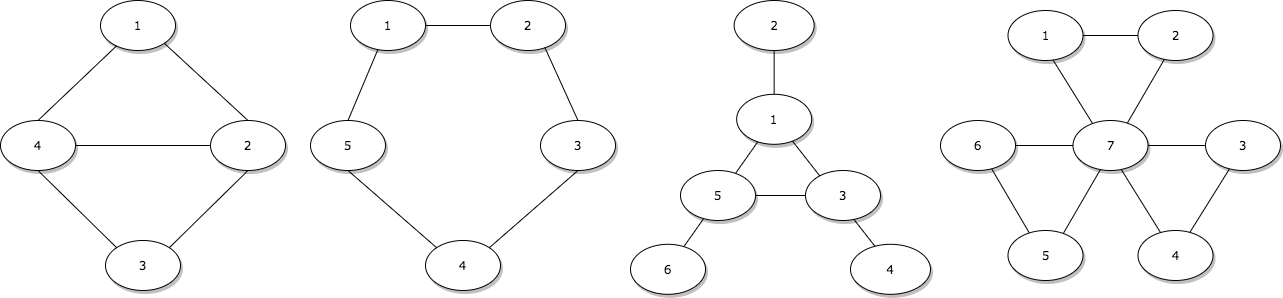}
\end{center}
\caption{Some examples where the global equivalence holds.}
\label{isingEg}
\end{figure}

Note that all these results are local statements with respect to a pair of nodes $(i,j)$ rather than global statements about the graphs. That is, for an interested pair $(i,j)$, as long as $|\mathcal{N}_i\setminus\{j\}|  \leqslant 2$ or $|\mathcal{N}_j\setminus\{i\}|  \leqslant 2$, the equivalence between CI and ACI will hold regardless of the other nodes. The local statements are turned into global statements if the node degree $|\mathcal{N}_i|  \leqslant 2$ for all $i\in\mathsf{V}$, in which case the entire MRF will be the same as DASG. Such graphs may consist of some loops and threads. Figure~\ref{isingEg} shows some special cases where global equivalence holds.

For example, consider the left most graph shown in Figure~\ref{isingEg}, where the node set is $\mathsf{V}=\{1,2,3,4\}$ and the edge set is $\mathcal{E} = \{(1,2),(2,3),(3,4),(1,4),(2,4)\}$. If the joint distribution is the symmetric Ising distribution specified by the first formula in \eqref{4nodegraph}, then the DAPO has blockwise norms given by the second formula in \eqref{4nodegraph}. Thus the graphs determined by ACI and CI are equivalent.

\begin{equation}\label{4nodegraph}
    \bm{\beta} = \frac{\log(2)}{2}\times\begin{pmatrix}0& 1& 0&1\\
1& 0&1&1\\
0& 1&0&1\\
1& 1&1&0
\end{pmatrix},
\quad\quad
     \Big\{\|\Theta_{X^iX^j}\|_\mathrm{HS}\Big\}_{i,j=1}^4 = \begin{pmatrix}\frac{11}{8}& \frac{33}{8}& 0& \frac{33}{8}\\
 \frac{33}{8}& \frac{1287}{800}& \frac{33}{8}& \frac{363}{800}\\
0& \frac{33}{8}& \frac{11}{8}& \frac{33}{8}\\
\frac{33}{8}& \frac{363}{800}& \frac{33}{8}& \frac{1287}{800}
\end{pmatrix}.
    \end{equation}

\begin{figure}[H]
  \begin{minipage}{.4\linewidth}
    \centering
\includegraphics[width=4cm]{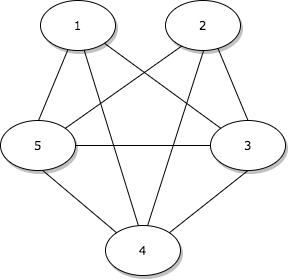}
  \end{minipage}
  \begin{minipage}{.4\linewidth}
    \centering
   \[\quad\bm{\beta} = \frac{\log(2)}{2}\times\begin{pmatrix}0&0&1& 1&1\\
0&0&1& 1&1\\
1&1&0& 1&1\\
1&1&1& 0&1\\
1&1&1& 1&0
\end{pmatrix}\]
  \end{minipage}
\caption{The fully connected symmetric Ising model except for the edge $(1,2)$. It has the unique $(1,2)$-separator $R=\mathcal{N}_1\setminus\{2\}=\mathcal{N}_2\setminus\{1\}=\{3,4,5\}$. In this case, $X^1\ind_A X^2 \mid  X^{\{3,4,5\}}$ is not true. The DASG yields a fully connected graph.}\label{counter}
\end{figure}

The condition $|\mathcal{N}_i\setminus\{j\}|  \leqslant 2$ or $|R| \leqslant 2$ requires a rather sparse edge set. The equivalence will not be guaranteed if they are violated. A simple counter example can be constructed when the graph has $5$ nodes and is fully connected except for one edge, as shown in Figure~\ref{counter}.

In this case, the blockwise norm of $\Theta_{XX}$ is given by
\begin{equation*}
\Big\{\|\Theta_{X^iX^j}\|_\mathrm{HS}\Big\}_{i,j=1}^5=\begin{pmatrix}\frac{10611}{5516} & \frac{27}{5516} & \frac{99}{197} & \frac{99}{197} & \frac{99}{197}\\
\frac{27}{5516} & \frac{10611}{5516} & \frac{99}{197} & \frac{99}{197} & \frac{99}{197}\\
\frac{99}{197} & \frac{99}{197} & \frac{474}{197} & \frac{117}{197} & \frac{117}{197}\\
\frac{99}{197} & \frac{99}{197} & \frac{117}{197} & \frac{474}{197} & \frac{117}{197}\\
\frac{99}{197} & \frac{99}{197} & \frac{117}{197} & \frac{117}{197} & \frac{474}{197}
\end{pmatrix}.
 \end{equation*}
The Hilbert-Schmidt norm of $\Theta_{X^1X^2}$ is $\|\Theta_{X^1X^2}\|_\mathrm{HS} = \frac{27}{5516}$, which is strictly greater than $0$. Unlike the conditional independence graph, the node variables here are all additive conditionally dependent on each other.

Next, consider the augmented graph defined by
$Y  \overset{\mathrm{d}}{=}  (X^1,X^2,X^3,X^4,X^5,X^3X^4X^5)^\top.$
In this case, it is easy to see that the MRF is the complete graph with the edge $(1,2)$ removed. The blockwise norms of $\Theta_{YY}$ is given by
\[\Big\{\|\Theta_{Y^iY^j}\|_\mathrm{HS}\Big\}_{i,j=1}^6=\begin{pmatrix}
 \frac{27}{14}& 0& \frac{15}{28}& \frac{15}{28}& \frac{15}{28}& \frac{3}{28}\\
0& \frac{27}{14}& \frac{15}{28}& \frac{15}{28}& \frac{15}{28}& \frac{3}{28}
\\
 \frac{15}{28}& \frac{15}{28}& \frac{221}{84}& \frac{31}{84}& \frac{31}{84}& \frac{61}{84}
\\
 \frac{15}{28}& \frac{15}{28}& \frac{31}{84}& \frac{221}{84}& \frac{31}{84}& \frac{61}{84}
\\
 \frac{15}{28}& \frac{15}{28}& \frac{31}{84}& \frac{31}{84}& \frac{221}{84}& \frac{61}{84}
\\
 \frac{3}{28}& \frac{3}{28}& \frac{61}{84}& \frac{61}{84}& \frac{61}{84}& \frac{197}{84}
\end{pmatrix}.\]
Thus the DASG agrees with the MRF again.
The key difference between $X$ and $Y$ lies in whether the linear conditional mean assumption \eqref{lcm} is satisfied. For $X$,
\[E[X^1\mid X^{\{3,4,5\}}] =E[X^2\mid X^{\{3,4,5\}}]= 1-\frac{2}{1+\exp[\log(2)(X^3+X^4+X^5)]},\]
neither of which is an element of $\mathscr{A}_{X^{\{3,4,5\}}}$. On the other hand, $(Y^3,Y^4,Y^5,Y^6)$ is actually a complete basis of odd function space of $Y^{\{3,4,5\}}$, which is slightly larger than $\mathscr{A}_{Y^{\{3,4,5\}}}$. The conditional expectation can then be expressed by:
\begin{equation*}
             E[Y^1\mid Y^{\{3,4,5,6\}}] =\frac{5}{18} (Y^3+Y^4+Y^5)-\frac{1}{18}Y^6.
\end{equation*}
That is to say, $Y^1$ has linear conditional mean with respect to $Y^{\{3,4,5,6\}}$.

\cite{loh2012structure} studied the relationship between CI and the inverse of a generalized covariance matrix by augmenting a graph with interaction terms. We exploit this idea to get a more complete picture about the relation between ACI and CI.
Let $D\subseteq  \mathsf{V}$ be a subset of nodes with $|D| \geqslant 3$. Let $$\mathscr{F}_D(X) := \Big\{\prod_{i\in D} u^{(n_i)}(X^i):n_i \in\{0,1\}\text{ and } \sum_{i\in D}n_i>1 \text{ and } \sum_{i\in D}n_i  = 1\ (\mathrm{mod}\ 2)\Big\},$$ where $u^{(0)}(x)=1$ and $u^{(1)}(x)=x$. Note that these terms are still binary. For example, if $D = \{3,4,5,6\}$,  $\mathscr{F}_D(X) = \{X^3X^4X^5,X^3X^4X^6,X^3X^5X^6,X^4X^5X^6\}$; if $D = \{3,4,5,6,7\}$,  the fifth order interaction $X^3X^4X^5X^6X^7$ is also contained in $\mathscr{F}_D(X)$.

Now, we may summarize the relation between ACI and CI in the next theorem.

\begin{theorem}\label{isingaugment}
Suppose $X = (X^1,X^2,\cdots,X^p)^\top$, $p\geqslant 5$, follows a symmetric Ising model \eqref{ising_sym} with respect to $\{\mathsf{V}, \mathcal{E}\}$. For any pair of nodes $(i,j)$ with $i\ne j$, let $R$ be a nonempty $(i,j)$-separator on $\{\mathsf{V}, \mathcal{E}\}$, and $Y$ be the augmented node vector with $Y^\top \overset{\mathrm{d}}{=} (X^\top,\mathscr{F}_{R}(X)^\top)$. Then
 with respect to $Y$,
$Y^i\ind Y^j \mid  Y^{-\{i,j\}}\Rightarrow  Y^i\ind_A Y^j \mid  Y^{-\{i,j\}}. $
Furthermore, if $R=\mathcal{N}_i\setminus \{j\}$, $Y^i\ind Y^j \mid  Y^{-\{i,j\}}\Leftrightarrow Y^i\ind_A Y^j \mid  Y^{-\{i,j\}}.$
\end{theorem}
This theorem provides some intuition of how much ACI and CI differ. The left hand side $Y^i\ind Y^j \mid  Y^{-\{i,j\}}$ is actually $X^i\ind X^j \mid  X^{-\{i,j\}}$ since we have the inclusion relationship between the $\sigma$-fields, $\sigma(\mathscr{F}_{R}(X))= \sigma(X^{R})\subseteq \sigma(X^{-\{i,j\}})$.
Note that Theorem \ref{isingaugment} is still a local result with respect to pair of nodes $(i,j)$ and the construction of $Y$ depends on $(i,j)$.

\section{Penalized estimation}\label{s4}
Given that $X$ follows a DASG with respect to $\mathcal{G} = \{\mathsf{V},\mathcal{E}\}$,
we need to estimate the edge set $\mathcal{E}$ from a sample of $X$. Since the relationship defined in Definition \ref{aci_def} is difficult to check directly, we look for a DAPO-based estimator.
We derive the matrix form of this operator in Subsection 4.1 and introduce a group penalized D-trace estimator in Subsection 4.2.

\subsection{Coordinate representation}
The discussion so far is in operator form. In this section, we represent the operators as matrices using coordinate representation. Two versions of such coordinate representations are derived, which are more direct than those given in \cite{li2014additive} thanks to the simplicity offered by the discrete $X$. We first give the matrix representations at the population level, which lead naturally to a sample estimate.

For each node variable $X^i$, consider its function space $\mathscr{A}_{X^i}=L^2(P_{X^i})$, the centered $L^2$ class. That is, for any $\phi\in \mathscr{A}_{X^i}$, $E\phi(X^i)=0$ and $E\phi(X^i)^2<\infty$. A natural way to represent such a function is by simply listing its values on the support, that is, $(\phi(0),\cdots,\phi(m))^\top$. Thus $\mathscr{A}_{X^i}$ has a finite dimension $\mathrm{dim}(\mathscr{A}_{X^i}) \leqslant m+1$.
Since the node variable $X^i$ takes its values in a finite set, $\phi(X^i)$ is a bounded random variable, whose second moment always exists. Therefore, the only restriction on $\phi$ is that its mean is zero and this renders the dimension of $\mathscr{A}_{X^i}$ as $m$.
We select an orthonormal basis $\big\{u_i^{(1)},\cdots,u_i^{(m)}\big\}$ of $\mathscr{A}_{X^i}$:
$\langle u_i^{(a)},u_i^{(b)}\rangle_{\mathscr{A}_{X^i}}=0$, $\langle u_i^{(a)},u_i^{(a)}\rangle_{\mathscr{A}_{X^i}}=1$, for any $a,b=1,\cdots,m \text{ and } a\ne b.$ Let $U_i(\cdot) =\big(
u_i^{(1)}(\cdot),\cdots,u_i^{(m)}(\cdot)
\big)^\top $. Then any function $\phi$ in $\mathscr{A}_{X^i}$ can be represented by a vector $c_\phi\in\mathbb{R}^{m}$ as $\phi(\cdot) = U_i(\cdot)^\top c_\phi$. With such a representation of functions, the $L^2(P_{X^i})$-inner product is just Euclidean inner product $\langle\phi, \psi\rangle_{\mathscr{A}_{X^i}} = c_\phi^\top c_\psi$. The function basis $U_i(\cdot)$, $i\in\mathsf{V}$, will give rise to the orthonormal representation of the DASG operator.

\begin{definition}(Orthonormal Representation)
 \begin{enumerate}[label=(\arabic*)]
    \item With $U_i(\cdot)$, $i\in \mathsf{V}$ defined above, the matrix $[\Sigma_{X^iX^j}]_{\mathrm{o}}:= \mathrm{cov}[U_i(X^i),U_j(X^j)]\in\mathbb{R}^{m\times m}$ is called an orthonormal representation of $\Sigma_{X^iX^j}$.
\item The matrix $[\Sigma_{XX}]_{\mathrm{o}} := \Big\{[\Sigma_{X^iX^j}]_{\mathrm{o}}\Big\}_{i,j=1}^p\in\mathbb{R}^{ mp\times mp}$ is called an orthonormal representation of the DAVO.
\item The matrix $[\Theta_{XX}]_{\mathrm{o}}:=[\Sigma_{XX}]_{\mathrm{o}}^{-1}\in\mathbb{R}^{ mp\times mp}$ is called an orthonormal representation of the DAPO. Let $[\Theta_{X^iX^j}]_{\mathrm{o}}\in \mathbb{R}^{ m\times m}$ denote the $(i,j)$-th block of $[\Theta_{XX}]_{\mathrm{o}}$.

\end{enumerate}
\end{definition}
The orthonormal representation is a  natural way to represent the operators. Note that ``orthonormal" refers to the function basis. Such a representation is not unique since it depends on the choice of $U_i(\cdot)$. If we are only interested in the signal (zero or nonzero) of a  block in the operators, there is a more convenient representation, as defined below.

\begin{definition}(Vertex Representation)
\begin{enumerate}[label=(\arabic*)]
    \item Let $V(\cdot) = \big(
\mathbbm{1}_{\{1\}}(\cdot),\cdots,\mathbbm{1}_{\{m\}}(\cdot)
\big)^\top$ be a vector-valued function on $\{0,1,\cdots, m\}$, whose $\ell$-th coordinate $\mathbbm{1}_{\{\ell\}}(\cdot)$ is the indicator function on $\{0,1,\cdots, m\}$. The matrix $[\Sigma_{X^iX^j}]_{\mathrm{v}}:= \mathrm{cov}[V(X^i),V(X^j)]\in\mathbb{R}^{ m\times m}$ is called the vertex representation of $\Sigma_{X^iX^j}$.

 \item The matrix $[\Sigma_{XX}]_{\mathrm{v}} := \Big\{[\Sigma_{X^iX^j}]_{\mathrm{v}}\Big\}_{i,j=1}^p\in\mathbb{R}^{ mp\times mp}$ is called the vertex representation of the DAVO.

 \item The matrix $[\Theta_{XX}]_{\mathrm{v}} := [\Sigma_{XX}]_{\mathrm{v}}^{-1}\in\mathbb{R}^{ mp\times mp}$ is called the vertex representation of the DAPO. Let $[\Theta_{X^iX^j}]_{\mathrm{v}}\in \mathbb{R}^{ m\times m}$ denote the $(i,j)$-th block of $[\Theta_{XX}]_{\mathrm{v}}$.

\end{enumerate}
\end{definition}
The word ``vertex" in the name comes from the fact that $V$ maps
$\{0,1,\cdots,m\}$ into $\{0,1\}^m$, which is the set of all vertices of the standard $m$-dimensional cube.
Replacing $U_i(\cdot)$'s with indicator function vector $V(\cdot)$ is computationally efficient. One may notice that $V(\cdot)$ itself is not a mean zero function basis. But it still works due to the fact that the constant functions vanish in the cross-covariance operator. The actual input function vector for node $i$ is $V(\cdot)-EV(X^i)$. Even though the constant functions are excluded for defining the operators, they can still be borrowed back for computational purposes.
\begin{theorem}\label{sgn}
The operators $\Sigma_{X^iX^j}$ and $\Theta_{X^iX^j}$ and their coordinate representations satisfy:
 \begin{enumerate}[label=(\arabic*)]
\item  $\Sigma_{X^iX^j}=0\Leftrightarrow [\Sigma_{X^iX^j}]_{\mathrm{o}}=0\Leftrightarrow[\Sigma_{X^iX^j}]_{\mathrm{v}}=0$.
\item  $\Theta_{X^iX^j}=0\Leftrightarrow [\Theta_{X^iX^j}]_{\mathrm{o}}=0\Leftrightarrow[\Theta_{X^iX^j}]_{\mathrm{v}}=0$.
\end{enumerate}
\end{theorem}
Within each statement, the first $0$ is the zero operator that maps any function in $\mathscr{A}_{X^j}$ to the zero function in $\mathscr{A}_{X^i}$. The second and the third $0$'s are $ m\times m$ zero matrices. We give two examples for coordinate representations and their roles in constructing the DASG.

\begin{example}Suppose
$X = (X^1,X^2,X^3)^\top\in\{0,1\}^3$ is a random vector with p.m.f. $f(x)=3^{x^1x^2}/12$.
The orthonormal and vertex representations of the DAVO and the DAPO are shown below.

  \begin{minipage}{.23\linewidth}
    \centering
   \[\begin{pmatrix}1& \frac{1}{4}& 0\\
 \frac{1}{4}& 1&0\\
0& 0&  1
\end{pmatrix}\]
$[\Sigma_{XX}]_{\mathrm{o}}$
  \end{minipage}%
  \begin{minipage}{.23\linewidth}
    \centering
  \[\begin{pmatrix}\frac{16}{15}& -\frac{4}{15}& 0\\
-\frac{4}{15}& \frac{16}{15}&0\\
0& 0&  1
\end{pmatrix}\]
$[\Theta_{XX}]_{\mathrm{o}}$
  \end{minipage}
   \begin{minipage}{.23\linewidth}
    \centering
   \[ \begin{pmatrix}\frac{2}{9}& \frac{1}{18}& 0\\
 \frac{1}{18}& \frac{2}{9}&0\\
0& 0&   \frac{1}{4}
\end{pmatrix}\]
$[\Sigma_{XX}]_{\mathrm{v}}$
  \end{minipage}%
  \begin{minipage}{.23\linewidth}
    \centering
  \[\begin{pmatrix}\frac{24}{5}& -\frac{6}{5}& 0\\
-\frac{6}{5}& \frac{24}{5}&0\\
0& 0&  4
\end{pmatrix}\]
$[\Theta_{XX}]_{\mathrm{v}}$
  \end{minipage}
\\\\
The matrix $[\Sigma_{XX}]_{\mathrm{o}}$ has $1$'s as its diagonal entries since the orthonormal basis $u_i$ actually normalizes the node variable. As a result, for a binary DASG, the edge set is determined by $[\Theta_{XX}]_{\mathrm{o}}=(\mathrm{corr}(X))^{-1}$, the inverse of the correlation matrix.
In comparison, it is easy to verify that the vertex representation $[\Sigma_{XX}]_{\mathrm{v}}$ is the covariance matrix $\mathrm{var}(X)$ and $[\Theta_{XX}]_{\mathrm{v}} = \left(\mathrm{var}(X)\right)^{-1}$. Clearly, both representations lead to the same graph.
\end{example}

\begin{example}
Suppose $X = (X^1,X^2,X^3)^\top\in\{0,1,2\}^3$ is a random vector with p.m.f. $f(x)=2^{x^1x^2+x^1x^3}/499$.
Take $V(\cdot)=(\mathbbm{1}_{\{1\}}(\cdot),\mathbbm{1}_{\{2\}}(\cdot))^\top$. For instance, the vertex representation of the cross-covariance operator $\Sigma_{X^1X^2}$ is
\[[\Sigma_{X^1X^2}]_{\mathrm{v}} =\mathrm{cov}[V(X^1),V(X^2)]=10^{-3}\times \begin{pmatrix}8.2&-16.1\\-10.5& 23.4 \end{pmatrix}. \]
The vertex representation of the DAPO is
\[[\Theta_{XX}]_{\mathrm{v}}=\begin{pmatrix}
[\Sigma_{X^1X^1}]_{\mathrm{v}}&[\Sigma_{X^1X^2}]_{\mathrm{v}}  & [\Sigma_{X^1X^3}]_{\mathrm{v}}     \\
[\Sigma_{X^2X^1}]_{\mathrm{v}} & [\Sigma_{X^2X^2}]_{\mathrm{v}} & [\Sigma_{X^2X^3}]_{\mathrm{v}}     \\
 [\Sigma_{X^3X^1}]_{\mathrm{v}}& [\Sigma_{X^3X^2}]_{\mathrm{v}}  & [\Sigma_{X^3X^3}]_{\mathrm{v}}     \\
\end{pmatrix}^{-1}=\begin{pmatrix}
67.0& 57.6& -2.9 &-3.5& -2.9 &-3.5    \\
57.6 &60.1& -4.1 &-5.4 &-4.1& -5.4   \\
 -2.9& -4.1& 21.4& 16.6&  0&  0  \\
 -3.5 &-5.4& 16.6& 18.2&  0&  0\\
 -2.9& -4.1&  0&  0& 21.4& 16.6\\
 -3.5& -5.4&  0&  0& 16.6& 18.2
\end{pmatrix}. \]
Note that the entire $(2,3)$-th block of $[\Theta_{XX}]_{\mathrm{v}}$ is zero, which means additive conditional independence holds between $X^2$ and $X^3$ given $X^1$.
\end{example}

We have assumed each node $X^i$ has the same support $\{0,\cdots,m\}$ for simplicity. But this is not necessary for constructing the DASG.
Suppose for
$X = (X^1,X^2,\cdots,X^p)^\top$, the node variable $X^i$ has the support $\{0,1,\cdots,m_i\}$ for $i\in\mathsf{V}$.
Let $V_i(\cdot) = (\mathbbm{1}_{\{1\}}(\cdot),\cdots,\mathbbm{1}_{\{m_i\}}(\cdot) )^\top$. The vertex representation of the DAVO is $[\Sigma_{XX}]_{\mathrm{v}}:=\Big\{[\Sigma_{X^iX^j}]_{\mathrm{v}}\Big\}_{i,j=1}^p$ with its $(i,j)$-th block being $[\Sigma_{X^iX^j}]_{\mathrm{v}}:= \mathrm{cov}[V_i(X^i),V_j(X^j)]\in\mathbb{R}^{m_i\times m_j}$. Hence, the off-diagonal sub-blocks may not be square matrices, as in the case of a common support $\{0,1,\cdots,m\}$.

\subsection{Group penalized D-trace estimator}
Suppose $X$ follows a DASG with respect to the graph $\mathcal{G} = \{\mathsf{V},\mathcal{E}\}$. We now develop an estimator of $\mathcal{E}$ based on an i.i.d. sample of $X$ of size $n$. In \cite{li2014additive} and \cite{lee2016additive}, $\mathcal{E}$ is estimated by thresholding the small entries of the estimated APO. Here, we propose a penalized sparse procedure to estimate it.

In the Gaussian graphical model setting, a well known method for estimating a positive definite sparse precision matrix is via the graphical Lasso or the $\ell_1$-penalized Gaussian likelihood estimator (\cite{yuan2007model} and \cite{friedman2008sparse}). This is similar to our problem, where estimating the edge set $\mathcal{E}$ reduces to sparse estimation
of $[\Theta_{XX}]_{\mathrm{v}}$, with its vanishing blocks determining the absence of edges. Instead of the negative Gaussian log-likelihood, we choose the loss function as the difference between two traces of operators (D-trace), $L_D(\Theta , \Sigma)=\frac12 \langle\Theta^2,\Sigma\rangle_\mathrm{F}-\mathrm{tr}(\Theta)$, proposed by \cite{zhang2014sparse}, where $\langle \cdot,\cdot\rangle_\mathrm{F}$ is the Frobenius inner product. This loss function is a smooth and convex function of $\Theta$ with a unique minimizer ${\Sigma}^{-1}$. The convexity ensures computational efficiency.

We first need an estimate of the coordinate representation of the DAVO. Since $[\Sigma_{XX}]_{\mathrm{v}}$ is the covariance matrix of $(V^\top(X^1),\cdots,V^\top(X^p))^\top$, its sample version becomes a natural choice. Let $[\hat{\Sigma}_{XX}]_{\mathrm{v}}:=\Big\{[\hat{\Sigma}_{X^iX^j}]_{\mathrm{v}}\Big\}_{i,j=1}^p$, whose $(i,j)$-th block is $[\hat{\Sigma}_{X^iX^j}]_{\mathrm{v}}:= \mathrm{cov}_n[V(X^i),V(X^j)]$, the sample covariance between $V(X^i)$ and $V(X^j)$. With $[\hat{\Sigma}_{XX}]_{\mathrm{v}}$ thus constructed, we propose to estimate the DAPO by
\begin{equation}\label{dlasso}
 [\hat{\Theta}_{XX}]_{\mathrm{v}} := \underset{\Theta = \Theta^\top}{\arg\min} \left\{ L_D(\Theta , [\hat{\Sigma}_{XX}]_{\mathrm{v}}) +\lambda_n P(\Theta)\right\},
\end{equation}
where $\lambda_n > 0$ is the regularization parameter, and $P(\Theta)$ is a penalty function, which is chosen to be $\sum_{1  \leqslant i,j  \leqslant p,\ i\ne j}\| \Theta_{[i,j]}\|_\mathrm{F}$ with $[i,j]$ representing the index set corresponding to the $(i,j)$-th block of a $p\times p$ block matrix. That is,
$[i,j]=\{(a,b)\in \mathbb{N}\times\mathbb{N}:  (i-1) m< a  \leqslant i m,\ \ (j-1) m< b  \leqslant j m\},$
where $\mathbb{N}=\{1,2,\cdots\}$ stands for the set of natural numbers. With the norm in the penalty being chosen as the Frobenius norm, $P(\Theta)$ is actually a group-Lasso penalty. By the definition of the orthonormal representaion $[\Theta_{XX}]_{\mathrm{o}}$, its Frobenius norm $\| [\Theta_{X^iX^j}]_{\mathrm{o}}\|_\mathrm{F}$ coincides with the Hilbert-Schmidt norm $\| \Theta_{X^iX^j}\|_\mathrm{HS}$. Here we use $\| [\Theta_{X^iX^j}]_{\mathrm{v}}\|_\mathrm{F}$ as a substitute for $\| [\Theta_{X^iX^j}]_{\mathrm{o}}\|_\mathrm{F}$ for computational efficiency. In Section \ref{s5}, we will provide a guarantee of the existence and uniqueness of the solution $ [\hat{\Theta}_{XX}]_{\mathrm{v}}$ with a sufficiently large $n$ and a proper choice of $\lambda_n$ even when $p>n$ causes the singularity issue of $[\hat{\Sigma}_{XX}]_{\mathrm{v}}$.

We use the alternating direction method of multipliers (ADMM) to solve the optimization problem. We introduce a new variable $\Theta_0$ that is a duplicate of $\Theta$, so that they deal with the loss and the penalty separately. The optimization problem is reformulated as
\[ \underset{\Theta = \Theta^\top,\ \Theta=\Theta_0 }{\arg\min}\Big\{\frac12 \langle\Theta^2,[\hat{\Sigma}_{XX}]_{\mathrm{v}}\rangle_\mathrm{F}-\mathrm{tr}(\Theta) +\lambda_n \sum_{i\ne j}\| (\Theta_0)_{[i,j]}\|_\mathrm{F}\Big\}.\]
The augmented Lagrangian is
\[\mathcal{L}(\Theta,\Theta_0,\Lambda)=\frac12 \langle\Theta^2,[\hat{\Sigma}_{XX}]_{\mathrm{v}}\rangle_\mathrm{F}-\mathrm{tr}(\Theta) +\lambda_n \sum_{i\ne j}\| (\Theta_0)_{[i,j]}\|_\mathrm{F}+ \langle\Lambda, \Theta-\Theta_0\rangle_\mathrm{F} + \frac{\rho}{2}\| \Theta-\Theta_0\|_\mathrm{F}^2, \]
where $\rho>0$ is a fixed number and $\Lambda$ is a matrix in $\mathbb{R}^{ mp\times mp}$.
We iteratively update the value of $(\Theta,\Theta_0,\Lambda)$. Given $(\Theta^{(t)},\Theta_0^{(t)},\Lambda^{(t)})$
at the $t$-th step, update the estimates by
\begin{align}
\Theta^{(t+1)} &=  \underset{\Theta = \Theta^\top}{\arg\min}\ \mathcal{L}(\Theta,\Theta_0^{(t)},\Lambda^{(t)}),\label{theta}\\
 \Theta^{(t+1)}_0 &=  \underset{\Theta_0}{\arg\min}\ \mathcal{L}(\Theta^{(t+1)},\Theta_0,\Lambda^{(t)}),\label{theta0}\\
\Lambda^{(t+1)}  &= \Lambda^{(t)}+\rho(\Theta^{(t+1)}-\Theta_0^{(t+1)}).\nonumber
\end{align}

For \eqref{theta}, the stepwise solution is
\[\Theta^{(t+1)} = H([\hat{\Sigma}_{XX}]_{\mathrm{v}}+\rho I_{mp}, I_{mp}+\rho\Theta^{(t)}_0-\Lambda^{(t)}),\]
where $I_{mp}\in \mathbb{R}^{ mp\times mp}$ is the identity matrix and the function $H(A,B)$ is defined by
$$H(A,B) :=  \underset{\Theta = \Theta^\top}{\arg\min}\Big\{\frac12\langle\Theta^2,A\rangle_\mathrm{F}-\langle \Theta,B\rangle_\mathrm{F}\Big\}.$$
If $A = D_A\Sigma_AD_A^\top$ is the eigenvalue decomposition of $A$, with ordered eigenvalues $\sigma_1 \geqslant\cdots \geqslant\sigma_{ mp}$, then $H(A,B)$ can be written down explicitly as
\[H(A,B) = D_A\{(D_A^\top BD_A)\circ C \}D_A^\top,\]
where $\circ$ denotes the Hadamard product and $C$ is the matrix $\Big\{\frac{2}{\sigma_a+\sigma_b}\Big\}_{a,b=1}^{mp}$.

For \eqref{theta0}, we have the stepwise solution
 \[\Theta_0^{(t+1)} = S(\Theta^{(t+1)}+\frac1\rho\Lambda^{(t)},\frac{\lambda_n}{\rho}),\]
where $S(A,\lambda)$ is the function
\[S(A,\lambda) :=\underset{\Theta_0 = \Theta_0^\top}{\arg\min}\Big\{\frac12\langle\Theta_0^2,I_{mp}\rangle_\mathrm{F}-\langle \Theta_0,A\rangle_\mathrm{F}+\lambda P(\Theta_0)\Big\}. \]
Given a symmetric matrix $A$, the $(i,j)$-th block of $S(A,\lambda)$ can be obtained by soft-thresholding:
\[S(A,\lambda)_{[i,j]}=\begin{cases}A_{[i,j]} & i=j\\
(1-\frac{\lambda\phantom{_\mathrm{F}}}{\| A_{[i,j]}\|_\mathrm{F}})A_{[i,j]} & i\ne j,\ \| A_{[i,j]}\|_\mathrm{F}>\lambda,\\
0& i\ne j,\ \| A_{[i,j]}\|_\mathrm{F}  \leqslant\lambda.
\end{cases}
\]

We summarize the procedure developed above as the following algorithm. For any matrix $M$ in $\mathbb{R}^{ mp\times mp}$, let $\mathrm{Diag}(M)$ denote its diagonal matrix, that is, $\mathrm{Diag}(M)=M\circ I_{mp}$.

\begin{algorithm}[H]
  \caption{D-trace group-Lasso}
  \begin{algorithmic}
\State Input: Sample DAVO $[\hat{\Sigma}_{XX}]_{\mathrm{v}}$, block size $ m$ and regularization parameter $\lambda_n$.
\State Set-up: Set proper $\rho$.

\State Initialize: Set $t=0$,  $\Theta^{(0)}=\Theta_0^{(0)}=[\mathrm{Diag}([\hat{\Sigma}_{XX}]_{\mathrm{v}})]^{-1}$ and $\Lambda^{(0)}=0$.

    \While {not converge}
      \State $\Theta^{(t+1)} = H([\hat{\Sigma}_{XX}]_{\mathrm{v}}+\rho I_{mp}, I_{mp}+\rho\Theta^{(t)}_0-\Lambda^{(t)})$.
      \State $\Theta_0^{(t+1)} = S(\Theta^{(t+1)}+\Lambda^{(t)}/\rho,\lambda_n/\rho)$.
      \State $\Lambda^{(t+1)} = \Lambda^{(t)}+\rho(\Theta^{(t+1)}-\Theta_0^{(t+1)})$.
      \State $t = t+1$.
    \EndWhile.
\State
Return: $\Theta_0^{(t)}$.
 \end{algorithmic}
\end{algorithm}
\section{Asymptotic properties}\label{s5}
In this section we derive the convergence rate of the penalized D-trace estimator \eqref{dlasso}. In the ultrahigh-dimensional setting, where $\log(p)$ is comparable to the sample size $n$, we will show that under an irrepresentable condition, the proposed estimator is consistent.

We first study the behavior of the sample-level coordinate representation of the DAVO. Since we use $[\hat{\Sigma}_{XX}]_{\mathrm{v}}$ to estimate $[\Sigma_{XX}]_{\mathrm{v}}$, the consistency requires a bound on
the difference $[\hat{\Sigma}_{XX}]_{\mathrm{v}}-[\Sigma_{XX}]_{\mathrm{v}}$ in the entrywise $\ell_\infty$-norm. For a matrix $M$, $\| M\|_{\infty}=\max_{ij}|M_{ij}|$.

\begin{lemma}\label{cov}
 For any constant $\tau>2$, with probability at least $1-1/p^{\tau-2}$, we have
\[\|[\hat{\Sigma}_{XX}]_{\mathrm{v}}-[\Sigma_{XX}]_{\mathrm{v}}\|_\infty  \leqslant \frac{3}{\sqrt{2}}\sqrt{\frac{\log(6 m^2)+\tau\log(p)}{n}}. \]
\end{lemma}
As with the classical Lasso, we need a type of irrepresentable condition for it to be consistent. For this purpose we introduce some additional
notations. Let $\mathcal{S}$ be the true blockwise support of $\Theta_{XX}$, that is $\mathcal{S}:= \big\{(i,j)\in\mathsf{V}\times \mathsf{V}: \Theta_{X^iX^j}\ne 0 \big\}.$
It is easy to see that $\mathcal{S}=\mathcal{E}\ \cup\ \{(i,i): i\in\mathsf{V}\}$, which is union of the edge set and the diagonal elements. Let $[i,j]$ be the index set defined in the previous section. Furthermore, for any subset
$A$ of $\mathsf{V}\times \mathsf{V}$, let $[A]:=\bigcup_{(i,j)\in A}[i,j]$ be the index set of the union of the blocks corresponding to the members of $A$.
Denoting the Kronecker product by $\otimes$, an important quantity involved in the irrepresentable condition is
\begin{equation}\label{Gamma}
\Gamma := \frac12\big([\Sigma_{XX}]_{\mathrm{v}}\otimes I_{mp}+I_{mp}\otimes[\Sigma_{XX}]_{\mathrm{v}}\big).
\end{equation}
For any two subsets
$B_1$ and $B_2$ of $\{1,\cdots, mp\} \times \{1,\cdots, mp\}$, let $\Gamma_{B_1,B_2}$ denote the submatrix of $\Gamma$ with rows and columns indexed by $B_1$ and $B_2$, that is, $\Gamma_{B_1,B_2}=\frac12\Big\{([\Sigma_{XX}]_{\mathrm{v}})_{ac}\delta_{bd}+([\Sigma_{XX}]_{\mathrm{v}})_{bd}\delta_{ac}\Big\}_{(a,b)\in B_1,\ (c,d)\in B_2},$
where $\delta_{\cdot\cdot}$ is the Kronecker delta function. The next assumption adapts the irrepresentable condition (\cite{ravikumar2011high}; \cite{zhang2014sparse}) to our blockwise setting. Note that, for any subset $A\subset\mathsf{V}\times \mathsf{V}$, $|[A]|=|A|m^2$.
\begin{assumption}
Let $\Upsilon :=\Gamma_{[\mathcal{S}^c],[\mathcal{S}]}(\Gamma_{[\mathcal{S}],[\mathcal{S}]})^{-1}\in  \mathbb{R}^{|\mathcal{S}^c| m^2\times |\mathcal{S}| m^2}$. Assume the following irrepresentable condition for the D-trace group-Lasso estimator \eqref{dlasso}:
\begin{equation}\label{irrep}
    \max_{e\in \mathcal{S}^c}  \sqrt{\sum_{f\in [e]}\Big( \sum_{e'\in \mathcal{S}}\| \Upsilon_{f,[e']}\|_2 \Big)^2}<1,
\end{equation}
where $\Upsilon_{f,[e']}$ is the $m^2$-dimensional vector $\{\Upsilon_{f,g}:g\in [e']\}$ and $\| \Upsilon_{f,[e']}\|_2$ is its Euclidean norm.
In the binary case with $m=1$, \eqref{irrep} reduces to
$\max_{e\in \mathcal{S}^c}  \sum_{e'\in \mathcal{S}}| \Upsilon_{e,e'}| <1. $
\end{assumption}
Let
$\gamma := 1- \max_{e\in \mathcal{S}^c}  \sqrt{\sum_{f\in [e]}\Big( \sum_{e'\in \mathcal{S}}\| \Upsilon_{f,[e']}\|_2 \Big)^2} ,$
then an equivalent way to express the irrepresentable condition is that $\gamma>0$. Assume the edge set $\mathcal{E}$ is sparse in the sense that the maximum degree is bounded, where the maximum degree is defined by
$d:= \max_{i\in\mathsf{V}}|\mathcal{N}_i|=\max_{i\in\mathsf{V}}\big| \big\{k\in\mathsf{V}\setminus\{i\}:(i,k)\in \mathcal{E}\big\}\big| ,$
which corresponds to its maximum number of nonzero blocks in any row of $[\Theta_{XX}]_{\mathrm{v}}$. For a matrix $M$, the norm $\| M\|_{1,\infty}$ means $\max_i(\sum_j |M_{ij}|)$. Let $\kappa_\Gamma := \|(\Gamma_{[\mathcal{S}],[\mathcal{S}]})^{-1}\|_{1,\infty}$ and $\kappa_\Sigma := \|[\Sigma_{XX}]_{\mathrm{v}}\|_{1,\infty}$. The three quantities, $d$, $\kappa_\Gamma$, and $\kappa_\Sigma$, control the degree of the sparsity of the graph.

We now establish the convergence rate under the assumption that the $n$ observations of $X$ are independently and identically sampled from a certain DASG.

\begin{theorem}\label{asymp}
Suppose $\kappa_\Gamma$, $\kappa_\Sigma$ and $d$ are bounded and the irrepresentable condition \eqref{irrep} holds. There exists $C>0$ such that for any constant $\tau>2$, if
$$\begin{cases}n>\max\big(m^{-2},\gamma^{-2}(\kappa_\Sigma\kappa_\Gamma+1)^2\big)\cdot \frac{81}{2} \kappa_\Gamma^2m^4d^2[\log(6 m^2)+\tau\log(p)]\\
\lambda_n=\frac{9}{\sqrt{2}}\gamma^{-1} (\kappa_\Sigma\kappa_\Gamma^2+\kappa_\Gamma)m^{\frac52}dn^{-\frac12}\sqrt{\log(6m^2)+\tau\log(p)}\end{cases},$$ then with probability at least $1-1/p^{\tau-2}$, the solution $[\hat{\Theta}_{XX}]_{\mathrm{v}}$ is unique and satisfies: \[\|[\hat{\Theta}_{XX}]_{\mathrm{v}}-[\Theta_{XX}]_{\mathrm{v}}\|_\infty\leqslant C  m^{\frac52}d\sqrt{\frac{\log(6m^2)+\tau\log(p)}{n}}.\]
\end{theorem}

The error bound of the binary DASG can be obtained as a special case. Assume the node vector $X$ is a member of $\{a_0,a_1\}^p$, where $a_0$ and $a_1$ are the two labels. The vertex representation of its DAPO is actually a scaled version of the inverse covariance matrix of $X$. For a matrix $M$, let $\| M\|_\mathrm{op}$ be its operator norm.
\begin{corollary}
The binary random vector $X$ follows the DASG with respect to $\mathcal{G} = \{\mathsf{V}, \mathcal{E}\}$. Suppose $\kappa_\Gamma$, $\kappa_\Sigma$ and $d$ are bounded and the irrepresentable condition \eqref{irrep} holds. There exists $C>0$ such that for any constant $\tau>2$, if
$$\begin{cases}n>\max\big(1,\gamma^{-2}(\kappa_\Sigma\kappa_\Gamma+1)^2\big)\cdot \frac{81}{2}\kappa_\Gamma^2d^2[\log(6 )+\tau\log(p)]\\
\lambda_n=\frac{9}{\sqrt{2}}\gamma^{-1}(\kappa_\Sigma\kappa_\Gamma^2+\kappa_\Gamma)dn^{-\frac12}\sqrt{\log(6)+\tau\log(p)}\end{cases},$$ then with probability at least $1-1/p^{\tau-2}$, the solution $[\hat{\Theta}_{XX}]_{\mathrm{v}}$ is unique and satisfies:
\begin{enumerate}[label=(\roman*)]
    \item $\|[\hat{\Theta}_{XX}]_{\mathrm{v}}-[\Theta_{XX}]_{\mathrm{v}}\|_\infty
  \ \leqslant C d\sqrt{\frac{\tau\log(p)}{n}},$
    \item $\|[\hat{\Theta}_{XX}]_{\mathrm{v}}-[\Theta_{XX}]_{\mathrm{v}}\|_{1,\infty}\leqslant C d^2\sqrt{\frac{\tau\log(p)}{n}},$
    \item $\|[\hat{\Theta}_{XX}]_{\mathrm{v}}-[\Theta_{XX}]_{\mathrm{v}}\|_\mathrm{op} \ \leqslant C\min\big(|\mathcal{S}|^{\frac12},d\big)  d\sqrt{\frac{\tau\log(p)}{n}}.$
\end{enumerate}
\end{corollary}

The above corollary shows that the convergence rate of our proposed estimator is the same as D-trace lasso estimator for Gaussian graphical model (\cite{zhang2014sparse}). It is also close to the graphical Lasso (\cite{ravikumar2011high}) up to the maximum node degree $d$. This is significant as our method is model free, whereas the graphical Lasso is a parametric method. Compared with the original APO-based estimator proposed by \cite{li2014additive} and \cite{lee2016additive}, whose optimal rate is $O(n^{-\frac14})$ (\cite{li2018nonparametric}), our new estimator substantially makes full use of the sparsity assumption. The improved rate also lies in the fact that there is no need to deal with nonparametric smoothing in the discrete setting.

\section{Simulation study}\label{s6}
In this section we compare our new D-trace group-Lasso DASG estimator (DLasso) with the APO-based estimator (APO; \cite{li2014additive} and \cite{lee2016additive}) and graphical Lasso (GLasso; \cite{friedman2008sparse}) for binary data. Since binary data does not follow the Gaussian model, the parametric assumption of GLasso is not satisfied. Nonetheless, its algorithm can be adapted to the blockwise setting to produce a DASG estimator. That is, the GLasso estimator in this section is
$$
 [\hat{\Theta}_{XX}]_{\mathrm{v}}^{(\mathrm{GLasso})} := \underset{\Theta = \Theta^\top}{\arg\min} \Big\{ \langle\Theta,[\hat{\Sigma}_{XX}]_{\mathrm{v}}\rangle_\mathrm{F}-
 \log\mathrm{det}(\Theta) +\lambda_n \sum_{i\ne j}\| \Theta_{[i,j]}\|_\mathrm{F}\Big\}.
$$
We also provide examples where the DASG and the MRF are equivalent, in which case the sparse Ising model (SpIsing; \cite{xue2012nonconcave}) is also included for comparison.

For any given edge set $\mathcal{E}$, we construct a random vector $X$ following a DASG with respect to $\mathcal{E}$ using the following procedure:
\begin{enumerate}
    \item  Choose a positive definite $A\in\mathbb{R}^{p\times p}$ such that $(i,j)\notin \mathcal{E} \Leftrightarrow A_{ij}=0$ and let $B = A^{-1}$.
    \item Let $\Sigma = C_B^{-\frac12}BC_B^{-\frac12}$, where $C_B=\mathrm{Diag}(B)$ is a diagonal matrix.
    \item Transform $\Sigma$ to $\Sigma'\in\mathbb{R}^{p\times p}$ by $\Sigma'_{ij}=\sin(\frac{\pi}{2}\Sigma_{ij})$. Suppose $\Sigma'$ is also positive definite.
    \item Generate a Gaussian random vector $W=(W^1,\cdots,W^p)^\top\sim N(0,\Sigma')$.
    \item Obtain the binary node vector $X=\mathrm{sign}(W) = (\mathrm{sign}(W^1),\cdots,\mathrm{sign}(W^p))^\top$.
\end{enumerate}
By construction, the DAPO for $X$ has the orthonormal representation as $[\Theta_{XX}]_{\mathrm{o}} = C_B^{\frac12}AC_B^{\frac12}$, which agrees with $\mathcal{E}$. The matrix $A$ in the first step is called a pattern matrix.

We consider the following models:
\begin{itemize}
    \item Model 1: Ising model with parameter
$\beta_{ij}= 0.3\cdot\mathbbm{1}_{\{|i-j|=1\}} + 0.3\cdot\mathbbm{1}_{\{|i-j|=p-1\}}$.
\item Model 2: Ising model with parameter
$\beta_{i,i+1}= 0.3$ for any $i\notin D$;
$\beta_{1j}= 0.2$ for $j\in D\setminus\{1\}$ or $(j-1)\in D$; and $\beta_{ij}= 0$ otherwise, where $D=\{1,p/50,2p/50,\cdots,49p/50,p\}$, with $p$ chosen to be $50q$ for some integer $q\geqslant 2$.

\item Model 3: binary DASG generated by the above procedure with the pattern matrix $A$, where
$A_{ij}= \mathbbm{1}_{\{i=j\}}+0.25\cdot\mathbbm{1}_{\{|i-j|=1\}}+0.15\cdot\mathbbm{1}_{\{|i-j|=2\}}$.

\item Model 4: binary DASG generated by the above procedure with the pattern matrix $A$, where
$A_{ij}= \mathbbm{1}_{\{i=j\}}+0.24\cdot0.75^{|i-j|-1}\cdot\mathbbm{1}_{\{1  \leqslant |i-j|  \leqslant 3\}}$.
\end{itemize}
We choose $n = 300$ and $p=200,\ 400$, and repeat the experiment $100$ times for each $(n,p)$. The DLasso and the GLasso are tuned by the five-fold cross validation and the APO is tuned by generalized cross validation.

\begin{table}
\caption{Comparison of true positive rate (TPR), true negative rate (TNR), and $F_1$ score when $(n,p)=(300,200)$. Reported numbers are averages over 100 independent runs, with standard errors given in parentheses.}
\begin{center}
    \begin{tabular}{lllll}
& Method & TPR(\%) & TNR(\%) &$F_1$ score(\%)\\ \hline
\multicolumn{1}{l|}{\multirow{4}{*}{Model 1}} & SpIsing   & $ 99.60 \ ( 0.49 )$ & $ 98.66 \ ( 0.27 )$ & $ 60.37 \ ( 4.72 )$  \\
\multicolumn{1}{l|}{} & APO  & $ 99.96 \ ( 0.14 )$ & $ 88.90 \ ( 0.19 )$ & $ 15.46 \ ( 0.22 )$ \\
\multicolumn{1}{l|}{} & GLasso  & $ 99.82 \ ( 0.42 )$ & $ 91.54 \ ( 3.35 )$ & $ 24.76 \ ( 17.81 )$    \\
\multicolumn{1}{l|}{} & DLasso  & $ 99.66 \ ( 0.41 )$ & $ 97.72 \ ( 0.22 )$ & $ 47.11 \ ( 2.41 )$  \\ \cline{2-5}
\multicolumn{1}{l|}{\multirow{4}{*}{Model 2}} & SpIsing   & $ 79.80 \ ( 2.05 )$ & $98.54 \ ( 0.30 )$ & $ 54.41 \ ( 3.91 )$  \\
\multicolumn{1}{l|}{} & APO  & $ 91.20 \ ( 4.81 )$ & $ 87.76 \ ( 0.28 )$ & $ 15.75 \ ( 0.62 )$  \\
\multicolumn{1}{l|}{} & GLasso  & $ 91.50 \ ( 5.84 )$ & $ 90.97 \ ( 3.57 )$ & $ 24.63 \ ( 14.72 )$    \\
\multicolumn{1}{l|}{} & DLasso  & $ 90.18 \ ( 3.20 )$ & $ 97.45 \ ( 0.26 )$ & $ 46.17 \ ( 2.10 )$  \\ \cline{2-5}
\multicolumn{1}{l|}{\multirow{3}{*}{Model 3}} & APO   & $ 86.45 \ ( 1.57 )$ & $ 85.09 \ ( 0.20 )$ & $ 18.82 \ ( 0.38 )$   \\
\multicolumn{1}{l|}{}  & GLasso  & $ 73.53 \ ( 9.57 )$ & $ 94.00 \ ( 5.58 )$ & $ 37.72 \ ( 10.80 )$\\
\multicolumn{1}{l|}{} & DLasso    & $ 74.84 \ ( 2.33 )$ & $ 97.02 \ ( 0.28 )$ & $ 46.64 \ ( 1.56 )$\\ \cline{2-5}

\multicolumn{1}{l|}{\multirow{3}{*}{Model 4}} & APO   & $ 76.20 \ ( 1.47 )$ & $ 86.13 \ ( 0.24 )$ & $ 24.31 \ ( 0.53 )$  \\
\multicolumn{1}{l|}{}& GLasso & $ 57.88 \ ( 8.25 )$ & $ 94.29 \ ( 3.54 )$ & $ 35.19 \ ( 4.37 )$ \\
\multicolumn{1}{l|}{}  & DLasso & $ 55.79 \ ( 3.24 )$ & $ 97.07 \ ( 0.39 )$ & $ 44.45 \ ( 1.14 )$
\end{tabular}
\end{center}
\label{tb1}
\end{table}

\begin{table}
\caption{Comparison of true positive rate (TPR), true negative rate (TNR), and $F_1$ score when $(n,p)=(300,400)$. Reported numbers are averages over 100 independent runs, with standard errors given in parentheses.}
\begin{center}
    \begin{tabular}{lllll}
 & Method & TPR(\%)              & TNR(\%) &$F_1$ score(\%)\\ \hline
\multicolumn{1}{l|}{\multirow{4}{*}{Model 1}} & SpIsing   & $ 99.31 \ ( 0.47 )$ & $99.32 \ ( 0.12 )$ & $ 59.63 \ ( 4.08 )$  \\
\multicolumn{1}{l|}{} & APO  & $ 99.95 \ ( 0.10 )$ & $ 89.32 \ ( 0.08 )$ & $ 8.62 \ ( 0.06 )$  \\
\multicolumn{1}{l|}{} & GLasso  & $ 99.53 \ ( 0.61 )$ & $ 95.26 \ ( 3.83 )$ & $ 32.40 \ ( 23.19 )$   \\
\multicolumn{1}{l|}{} & DLasso  & $ 99.36 \ ( 0.42 )$ & $ 98.87 \ ( 0.09 )$ & $ 46.99 \ ( 1.90 )$  \\ \cline{2-5} \multicolumn{1}{l|}{\multirow{4}{*}{Model 2}} & SpIsing   & $ 87.30 \ ( 1.11 )$ & $99.34 \ ( 0.11 )$ & $ 57.63 \ ( 3.38 )$  \\
\multicolumn{1}{l|}{} & APO  & $94.83 \ ( 2.29 )$ & $ 90.10 \ ( 0.11 )$ & $ 9.76 \ ( 0.22 )$  \\
\multicolumn{1}{l|}{} & GLasso  & $ 92.76 \ ( 3.93 )$ & $ 95.95 \ ( 3.76 )$ & $ 36.06 \ ( 21.99 )$    \\
\multicolumn{1}{l|}{} & DLasso  & $ 92.51 \ ( 1.78 )$ & $ 98.85 \ ( 0.09 )$ & $ 46.90 \ ( 1.67 )$  \\ \cline{2-5}
\multicolumn{1}{l|}{\multirow{3}{*}{Model 3}} & APO   & $ 83.53 \ ( 1.22 )$ & $ 88.17 \ ( 0.08 )$ & $ 12.32 \ ( 0.18 )$   \\
\multicolumn{1}{l|}{}  & GLasso  & $ 64.99 \ ( 4.67 )$ & $ 97.66 \ ( 0.72 )$ & $ 33.55 \ ( 4.52 )$\\
\multicolumn{1}{l|}{}                         & DLasso    & $64.00\ ( 1.77 )$    & $98.74\ ( 0.09 )$ &$44.32\ ( 1.31 )$  \\ \cline{2-5}

\multicolumn{1}{l|}{\multirow{3}{*}{Model 4}} & APO   &$ 71.55 \ ( 1.01 )$ & $ 88.71 \ ( 0.09 )$ & $ 15.65 \ ( 0.24 )$  \\
\multicolumn{1}{l|}{}  & GLasso  & $ 49.61 \ ( 11.33 )$ & $ 96.37 \ ( 1.54 )$ & $ 26.03 \ ( 2.92 )$\\
\multicolumn{1}{l|}{}  & DLasso    & $
44.34\ (1.31)$    & $98.76\ ( 0.09 )$ &$39.23\ ( 1.06 )$
\end{tabular}
\end{center}
\label{tb2}
\end{table}

 Let $\hat{\mathcal{E}}$ denote the estimated edge set. We use three criteria to compare the prediction performance: the true positive rate (TPR), the true negative rate (TNR), and the $F_1$ score, as defined below:
$\mathrm{TPR} = \frac{|\hat{\mathcal{E}}\cap\mathcal{E}|}{|\mathcal{E}|},\ \mathrm{TNR} = \frac{|\hat{\mathcal{E}}^c\cap\mathcal{E}^c|}{|\mathcal{E}^c|},\ F_1=\frac{2|\hat{\mathcal{E}}\cap\mathcal{E}|}{|\mathcal{E}|+|\hat{\mathcal{E}}|}.$
TPR is the proportion of correctly estimated nonzeros; TNR is the proportion of correctly estimated zeros; $F_1$ score is the harmonic mean of $|\hat{\mathcal{E}}\cap\mathcal{E}|/|\hat{\mathcal{E}}|$ and $|\hat{\mathcal{E}}\cap\mathcal{E}|/|\mathcal{E}|$.

The simulation results are shown in Table~\ref{tb1} ($p=200$) and Table~\ref{tb2} ($p=400$). We see that DLasso has good accuracy. For Model 1, DLasso, as a model-free estimator, yields comparable results to the parametric SpIsing with close TPR and TNR. For Model 2, the DASG estimators give higher TPR than SpIsing, and DLasso maintains high TNR as well. For Models 3 and 4, DLasso achieves the highest TNR and $F_1$ score, and an acceptably high TPR. APO, the norm thresholding estimator, tends to give the highest TPR at the cost of the lowest TNR. GLasso behaves similarly to DLasso, but its TNR is uniformly lower than DLasso. By taking advantage of the sparsity of the graph, DLasso uncovers significantly more true negatives (true absent edges) than GLasso. On the other hand, DLasso gives much smaller standard errors in TPR and TNR than those of GLasso, reflecting its robustness due to the D-trace loss function.

\begin{figure}[!htp]
\begin{center}
\begin{subfigure}{.4\textwidth}
\includegraphics[width=\textwidth]{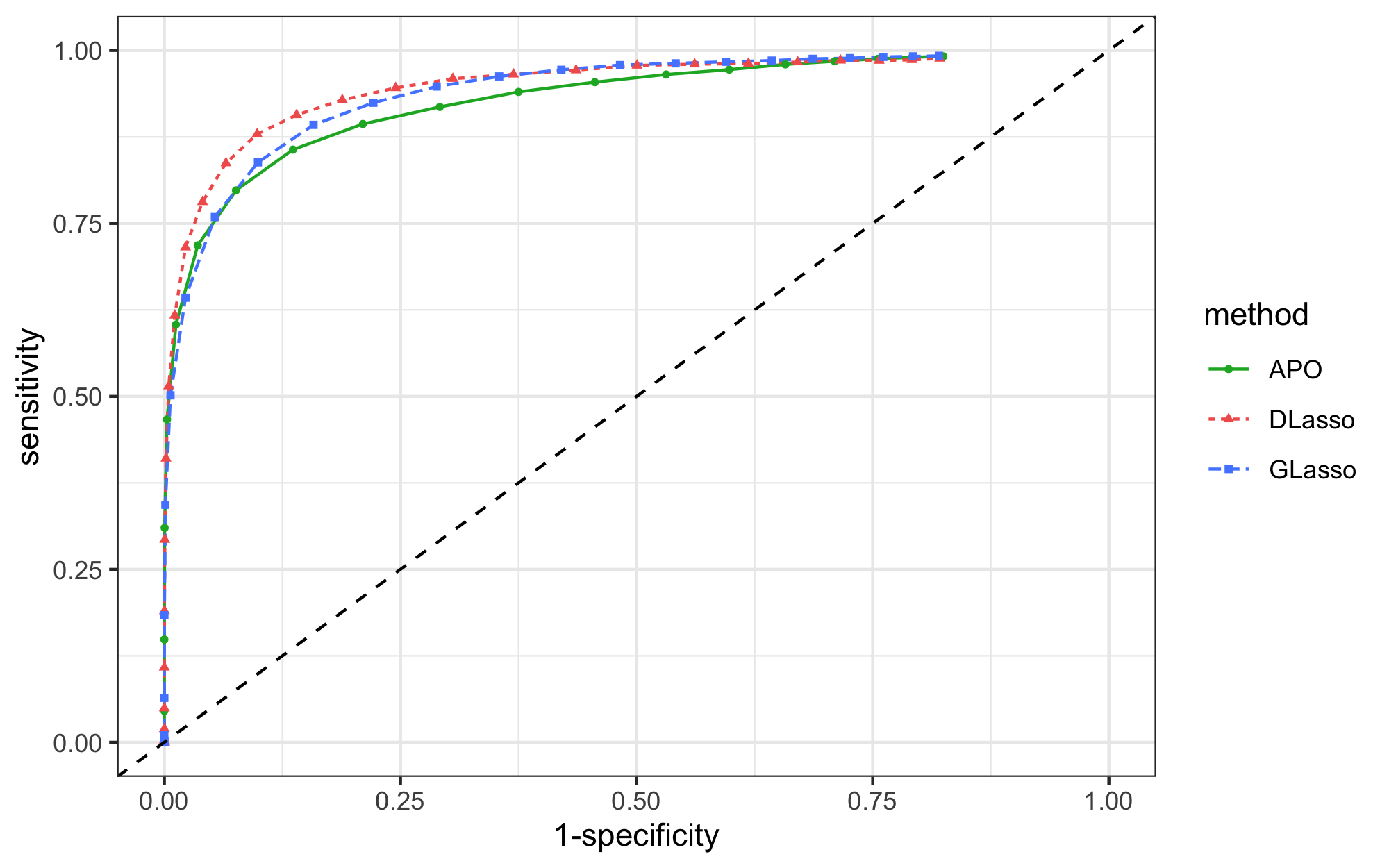}
\caption{Model 3 with $p=200$.}
\end{subfigure}
\begin{subfigure}{.4\textwidth}
\includegraphics[width=\textwidth]{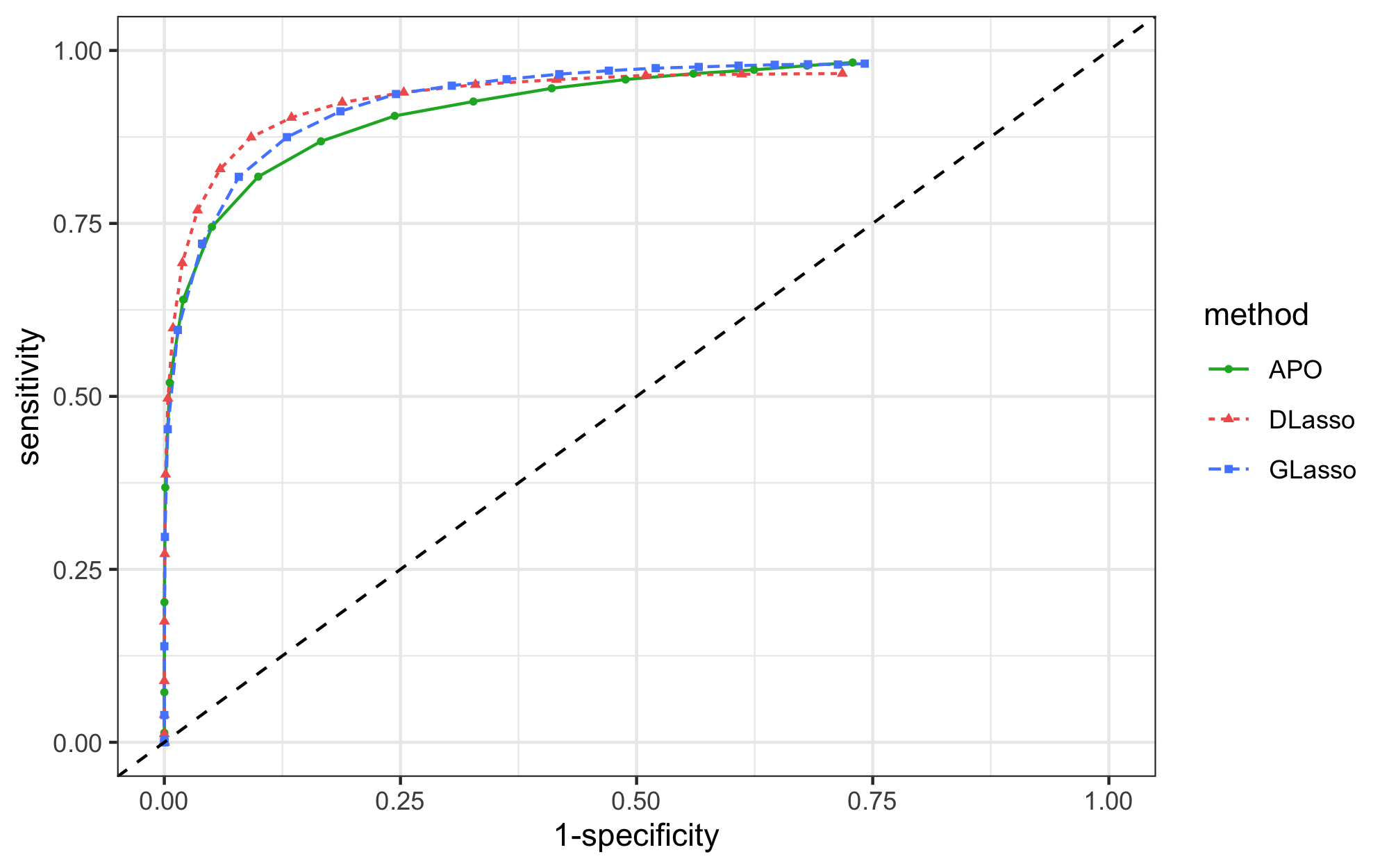}
\caption{Model 3 with $p=400$.}
\end{subfigure}
\\
\begin{subfigure}{.4\textwidth}
\includegraphics[width=\textwidth]{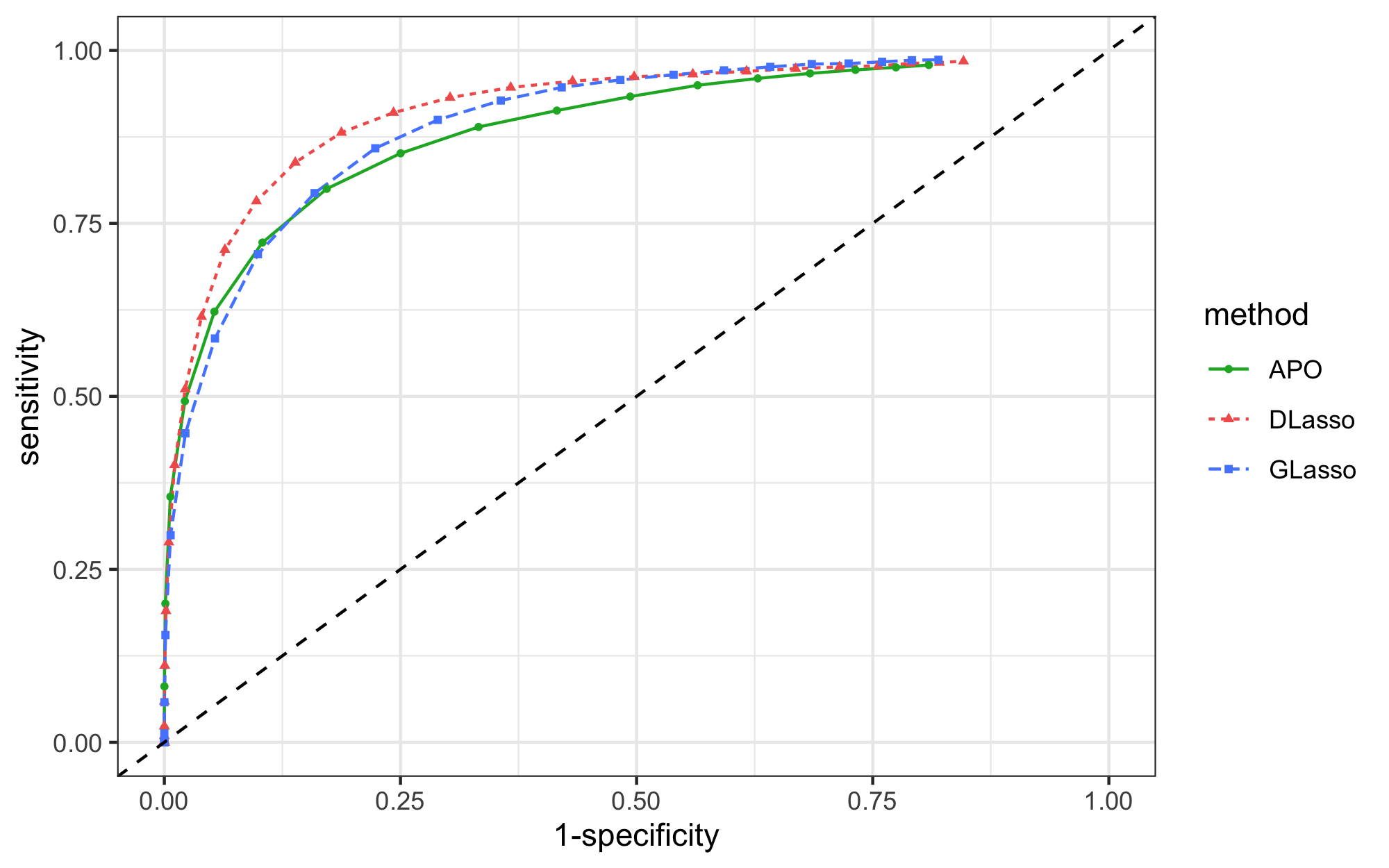}
\caption{Model 4 with $p=200$.}
\end{subfigure}
\begin{subfigure}{.4\textwidth}
\includegraphics[width=\textwidth]{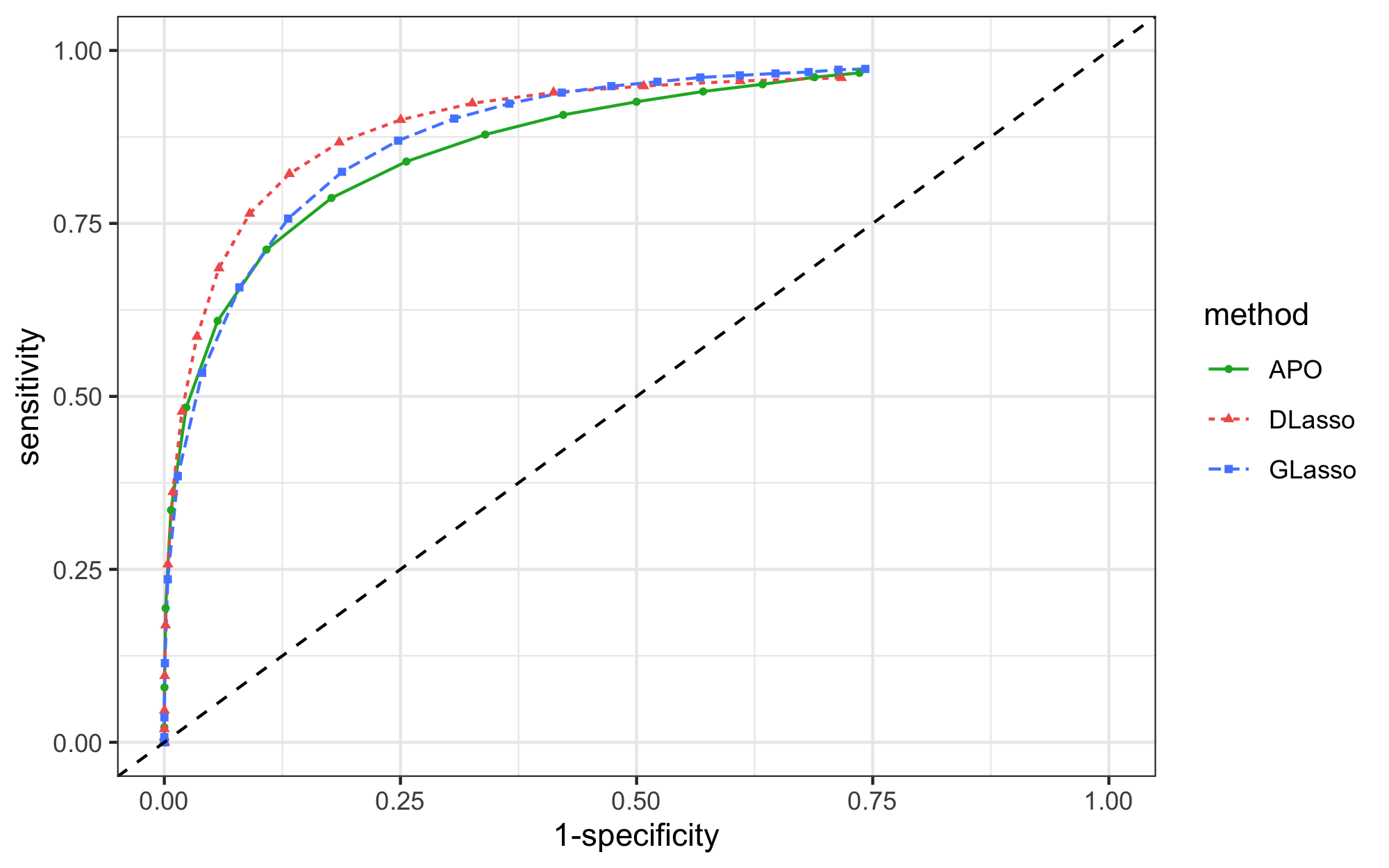}
\caption{Model 4 with $p=400$.}
\end{subfigure}
\end{center}
\caption{The comparison of ROC curves with respect to binary DASG's. }
\label{roc2}
\end{figure}

We also computed the receiver operating characteristic curves (ROC curves) of the estimators, which are presented in Figure~\ref{roc2}. Each curve is the average of ROC curves across $10$ simulation runs. For Models 3 and 4, although the GLasso curve is higher than the DLasso curve on the right end of the plot, the areas under curve (AUC) for the DLasso are substantially larger than both GLasso and APO.

\section{Application to HIV antiretroviral therapy data}\label{s7}
High throughput technology in biomedicine produces enormous data, offering an integrated view of life for scientists and clinicians. Processing and interpreting these data are in the frontiers of statistical methodological research. By presenting pairwise connections between variables intuitively, graphical models have been widely applied in disease diagnosis, drug discovery and prediction of regulation networks. In this section, the DASG is utilized to detect association between mutations in a data set of HIV-1 protease sequence. It could provide evidence for drug discovery in the future.

We apply our method to an HIV antiretroviral therapy (ART) susceptibility data set, which is a binary data set, as described in \cite{rhee2006genotypic}. The data set includes virus mutation information at $99$ protease residues for $n = 702$ isolates from the plasma of HIV-1-infected patients. A mutation for a certain protease residue is denoted by $+1$, and no mutation is denoted by $-1$. Our analysis only includes $p = 62$ of the $99$ residues that contain at least $5$ mutations.
\begin{figure}
\begin{center}
\includegraphics[width=9cm]{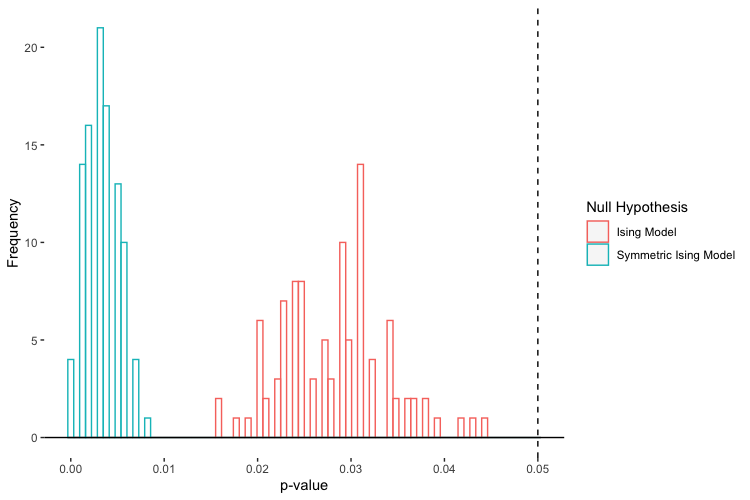}
\end{center}
\caption{Histograms of the $p$-values of the KDSD-tests for  Ising models.}
\label{kdsd}
\end{figure}

The CI graphical structure of this data set was studied by \cite{xue2012nonconcave} using the Ising model based method. However, it is questionable whether the data really follows an Ising model. Recent work by \cite{yang2018goodness} provided a goodness-of-fit test for the Ising model assumption via the Kernelized Discrete Stein Discrepancy (KDSD). We perform the KDSD test for two null hypotheses, the Ising model \eqref{ising_def} and the symmetric Ising model \eqref{ising_sym}. Since the KDSD-test is a bootstrap-based test, we repeat it $100$ times. For the Ising model \eqref{ising_def} hypothesis, we obtain an average p-value of $0.0281$ and the maximum p-value of $0.044$; for the symmetric Ising model \eqref{ising_sym} hypothesis, this average p-value is $0.0034$ and the maximum p-value is $0.008$. The histograms for the two sets of p-values for the two hypotheses are presented in Figure~\ref{kdsd}. Both the Ising model hypothesis and the symmetric Ising model hypothesis are rejected with significance level $0.05$. Especially, for the symmetric Ising model, the p-value is less than $0.01$. This can be partly explained by the mean values of the data, as zero expectation is a basic property satisfied by the symmetric Ising model \eqref{ising_sym}. In fact, except for residues $10$, $63$, and $71$, the mutation frequency at all the other residues are lower than $0.38$, leading to mostly negative column mean values. For that reason, a symmetric Ising model is not a good choice for CI structure learning.

\begin{figure}
  \centering
\begin{subfigure}{.4\textwidth}
  \centering
  \includegraphics[width=\textwidth]{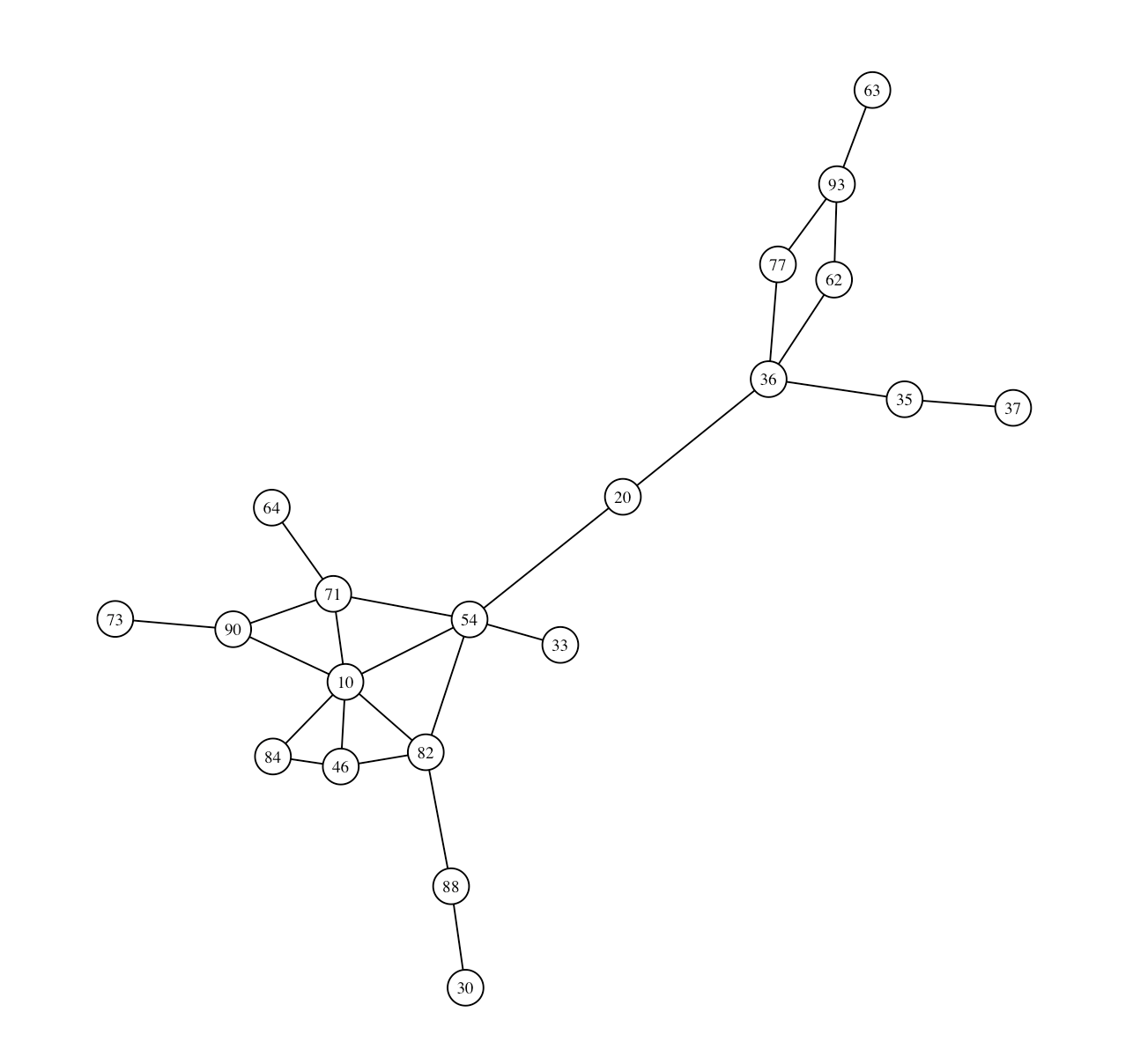}
  \caption{Stable edges of the Ising model.}
\end{subfigure}\quad
\begin{subfigure}{.4\textwidth}
  \centering
  \includegraphics[width=\textwidth]{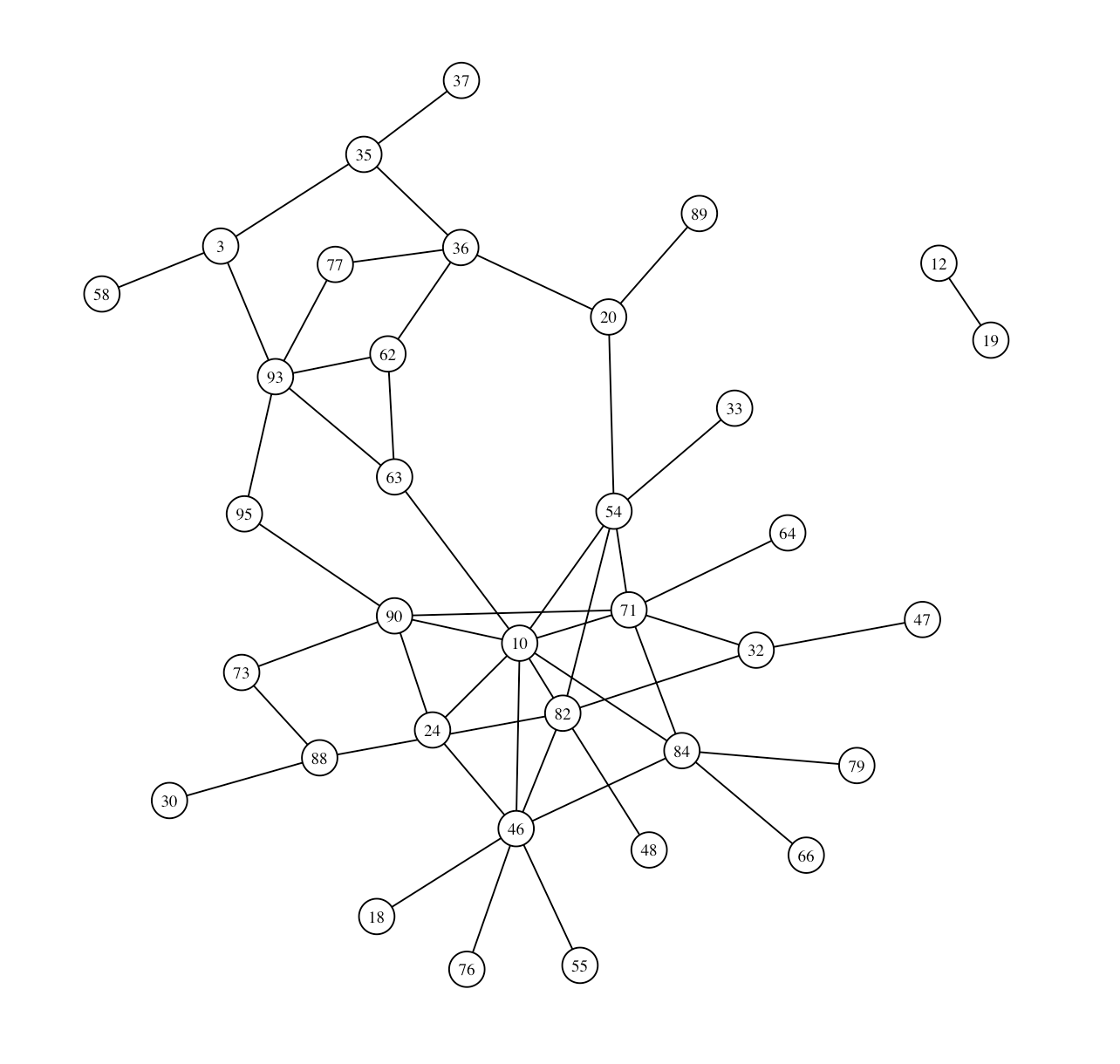}
  \caption{Stable edges of the DASG.}
  \label{fig:sub-second}
\end{subfigure}
\caption{Comparison of the Ising model (MRF) and the DASG on the HIV data. 
}
 \label{fig:hiv_asg}
\end{figure}

Thus, by the test results, the Ising model does not seem to be a valid assumption. It is then reasonable to turn to a nonparametric method such as the DASG.
In Figure~\ref{fig:hiv_asg} we present the bootstrapped stable edges of the Ising model and the DASG: we draw $100$ bootstrap samples of size $702$ and estimate the Ising model and the DASG repeatedly, and plot the edges selected more than $95$ times.

In the following discussion, a string such as ``x", ``Ax", and ``AxB", where ``A" and ``B" are capital letters and ``x" is an integer, corresponds to a node represented by the number ``x" in the graphs in Figure~\ref{fig:hiv_asg}. With ``x" standing for residue x, ``Ax" means that its original type of amino acid is A and ``AxB" means the mutation into type B amino acid.

The stable edge set of the DASG includes $48$ edges. Most of them are meaningful in the context of HIV study, which can be verified by some previous literature. \cite{hoffman2003covariation} discovered and explained many pairs of interactions which are consistent with our findings. For example, in the K20R:M36I double mutant, where K20 and M36 are physically close to each other, the double mutation would compensate for the original interaction between the two residues (\cite{hoffman2003covariation}); Residues 35 and 37 are near one
another in the protease structure in the hinge region of the flap. E35 forms an ionic bond with R57 and N37D would be stabilized by interacting with R57 (\cite{hoffman2003covariation}); An M46I mutant exhibits enhanced catalytic activity over the wild-type enzyme and improves the activity of a protease mutant containing both V82T and I84V (\cite{schock1996mutational} and \cite{hoffman2003covariation}); The double mutation D30N and N88D can reduce nelfinavir susceptibility by 50-fold (\cite{rhee2003human} and \cite{liu2008analysis}).

\begin{figure}
\begin{center}
\includegraphics[width=9cm]{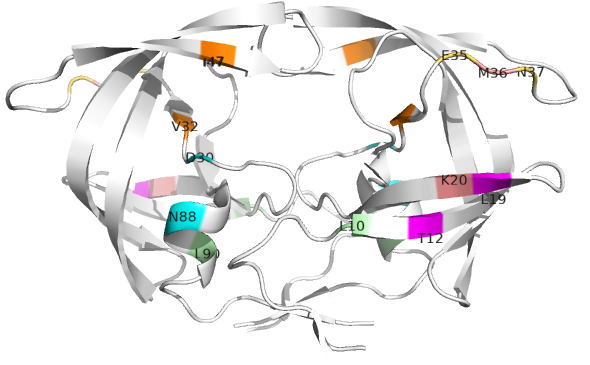}
\end{center}
\caption{A ribbon diagram of the HIV-1 protease dimer about the locations of some double mutations found by the DASG, with each pair of residues highlighted in the same color. The diagram was generated from the  molecular visualization system PyMOL (\cite{delano2002pymol}).}
\label{hiv}
\end{figure}
The DASG identifies residue $10$ as an important node hub. L10I, although not causing resistance alone (it is a minor resistance residue), plays a critical
role in eliciting the cooperative response along with L90M at the dimer interface
(\cite{ohtaka2003multidrug} and \cite{liu2008analysis}). Mutations at both residues 10 and 71 frequently appear in clinical samples from subjects who have failed protease inhibitor therapy and these mutations can be selected by passage of HIV-1 in the presence of a protease inhibitor in vitro (\cite{hoffman2003covariation}).

Compared with the DASG, the Ising model recovers only the most significant pairs with $25$ stable edges. Based on $2,244$ subtype B HIV-1 isolates from $1,919$ persons with different protease inhibitor experiences, \cite{wu2003mutation}
reported $(54, 82)$ and $(32, 47)$ as two of the most highly correlated pairs. \cite{hoffman2003covariation} pointed out that residues 48, 54, and 82 represent a hierarchy of interactions, where G48V is likely a late mutation added after changes at both 54 and 82 have occurred. The Ising model finds $(54, 82)$ but misses their interactions with 48; HIV-1 protease with drug resistant mutations V32I, I47V and V82I has been evaluated as a model for inhibition of HIV-2 protease to overcome the problem of autoproteolysis of HIV-2 protease. There are hydrophobic interactions of residues 32 and 47 according to molecular biology analysis (\cite{pawar2019structural}). The Ising model fails to keep $(32,47)$ as an important edge while the DASG captures it successfully. Another pair uncovered by the DASG only is $(12,19)$, and the interaction between residues 12 and 19 seems to be necessary to maintain Van der Waals force (\cite{hoffman2003covariation}).


\bibliographystyle{agsm}
\bibliography{graph}

@article{guo2015graphical,
  title={Graphical models for ordinal data},
  author={Guo, Jian and Levina, Elizaveta and Michailidis, George and Zhu, Ji},
  journal={Journal of Computational and Graphical Statistics},
  volume={24},
  number={1},
  pages={183--204},
  year={2015},
  publisher={Taylor \& Francis}
}

@article{lee2021estimating,
  title={Estimating Finite Mixtures of Ordinal Graphical Models},
  author={Lee, Kevin H and Chen, Qian and DeSarbo, Wayne S and Xue, Lingzhou},
  journal={Psychometrika, in press},
  volume={},
  pages={},
  year={2021},
  publisher={Springer}
}

@article{ising1925beitrag,
  title={Beitrag zur theorie des ferromagnetismus},
  author={Ising, Ernst},
  journal={Zeitschrift f{\"u}r Physik},
  volume={31},
  number={1},
  pages={253--258},
  year={1925},
  publisher={Springer}
}

@article{li2014additive,
  title={On an additive semigraphoid model for statistical networks with application to pathway analysis},
  author={Li, Bing and Chun, Hyonho and Zhao, Hongyu},
  journal={Journal of the American Statistical Association},
  volume={109},
  number={507},
  pages={1188--1204},
  year={2014},
  publisher={Taylor \& Francis}
}

@article{li2018nonparametric,
  title={A nonparametric graphical model for functional data with application to brain networks based on fMRI},
  author={Li, Bing and Solea, Eftychia},
  journal={Journal of the American Statistical Association},
  volume={113},
  number={524},
  pages={1637--1655},
  year={2018},
  publisher={Taylor \& Francis}
}

@article{zhang2014sparse,
  title={Sparse precision matrix estimation via lasso penalized D-trace loss},
  author={Zhang, Teng and Zou, Hui},
  journal={Biometrika},
  volume={101},
  number={1},
  pages={103--120},
  year={2014},
  publisher={Oxford University Press}
}

@inproceedings{loh2012structure,
  title={Structure estimation for discrete graphical models: Generalized covariance matrices and their inverses},
  author={Loh, Po-Ling and Wainwright, Martin J},
  booktitle={Advances in Neural Information Processing Systems},
  pages={2087--2095},
  year={2012}
}

@book{pearl1987logic,
  title={The Logic of Representing Dependencies by Directed Graphs},
  author={Pearl, Judea and Verma, Thomas},
  year={1987},
  publisher={University of California (Los Angeles). Computer Science Department}
}

@article{baker1973joint,
  title={Joint measures and cross-covariance operators},
  author={Baker, Charles R},
  journal={Transactions of the American Mathematical Society},
  volume={186},
  pages={273--289},
  year={1973}
}

@article{fukumizu2009kernel,
  title={Kernel dimension reduction in regression},
  author={Fukumizu, Kenji and Bach, Francis R and Jordan, Michael I},
  journal={The Annals of Statistics},
  volume={37},
  number={4},
  pages={1871--1905},
  year={2009},
  publisher={Institute of Mathematical Statistics}
}

@article{xue2012nonconcave,
  title={Nonconcave penalized composite conditional likelihood estimation of sparse Ising models},
  author={Xue, Lingzhou and Zou, Hui and Cai, Tianxi},
  journal={The Annals of Statistics},
  volume={40},
  number={3},
  pages={1403--1429},
  year={2012},
  publisher={Institute of Mathematical Statistics}
}

@inproceedings{yang2018goodness,
  title={Goodness-of-fit testing for discrete distributions via Stein discrepancy},
  author={Yang, Jiasen and Liu, Qiang and Rao, Vinayak and Neville, Jennifer},
  booktitle={International Conference on Machine Learning},
  pages={5561--5570},
  year={2018}
}

@article{rhee2006genotypic,
  title={Genotypic predictors of human immunodeficiency virus type 1 drug resistance},
  author={Rhee, Soo-Yon and Taylor, Jonathan and Wadhera, Gauhar and Ben-Hur, Asa and Brutlag, Douglas L and Shafer, Robert W},
  journal={Proceedings of the National Academy of Sciences},
  volume={103},
  number={46},
  pages={17355--17360},
  year={2006},
  publisher={National Acad Sciences}
}

@article{wu2003mutation,
  title={Mutation patterns and structural correlates in human immunodeficiency virus type 1 protease following different protease inhibitor treatments},
  author={Wu, Thomas D and Schiffer, Celia A and Gonzales, Matthew J and Taylor, Jonathan and Kantor, Rami and Chou, Sunwen and Israelski, Dennis and Zolopa, Andrew R and Fessel, W Jeffrey and Shafer, Robert W},
  journal={Journal of Virology},
  volume={77},
  number={8},
  pages={4836--4847},
  year={2003},
  publisher={Am Soc Microbiol}
}

@article{ohtaka2003multidrug,
  title={Multidrug resistance to HIV-1 protease inhibition requires cooperative coupling between distal mutations},
  author={Ohtaka, Hiroyasu and Sch{\"o}n, Arne and Freire, Ernesto},
  journal={Biochemistry},
  volume={42},
  number={46},
  pages={13659--13666},
  year={2003},
  publisher={ACS Publications}
}

@article{liu2008analysis,
  title={Analysis of correlated mutations in HIV-1 protease using spectral clustering},
  author={Liu, Ying and Eyal, Eran and Bahar, Ivet},
  journal={Bioinformatics},
  volume={24},
  number={10},
  pages={1243--1250},
  year={2008},
  publisher={Oxford University Press}
}

@article{hoffman2003covariation,
  title={Covariation of amino acid positions in HIV-1 protease},
  author={Hoffman, Noah G and Schiffer, Celia A and Swanstrom, Ronald},
  journal={Virology},
  volume={314},
  number={2},
  pages={536--548},
  year={2003},
  publisher={Elsevier}
}

@article{pawar2019structural,
  title={Structural studies of antiviral inhibitor with HIV-1 protease bearing drug resistant substitutions of V32I, I47V and V82I},
  author={Pawar, Shrikant and Wang, Yuan-Fang and Wong-Sam, Andres and Agniswamy, Johnson and Ghosh, Arun K and Harrison, Robert W and Weber, Irene T},
  journal={Biochemical and Biophysical Research Communications},
  volume={514},
  number={3},
  pages={974--978},
  year={2019},
  publisher={Elsevier}
}

@article{schock1996mutational,
  title={Mutational anatomy of an HIV-1 protease variant conferring cross-resistance to protease inhibitors in clinical trials compensatory modulations of binding and activity},
  author={Schock, Hilary B and Garsky, Victor M and Kuo, Lawrence C},
  journal={Journal of Biological Chemistry},
  volume={271},
  number={50},
  pages={31957--31963},
  year={1996},
  publisher={ASBMB}
}

@article{rhee2003human,
  title={Human immunodeficiency virus reverse transcriptase and protease sequence database},
  author={Rhee, Soo-Yon and Gonzales, Matthew J and Kantor, Rami and Betts, Bradley J and Ravela, Jaideep and Shafer, Robert W},
  journal={Nucleic Acids Research},
  volume={31},
  number={1},
  pages={298--303},
  year={2003},
  publisher={Oxford University Press}
}

@article{lee2016additive,
  title={On an additive partial correlation operator and nonparametric estimation of graphical models},
  author={Lee, Kuang-Yao and Li, Bing and Zhao, Hongyu},
  journal={Biometrika},
  volume={103},
  number={3},
  pages={513--530},
  year={2016},
  publisher={Oxford University Press}
}

@article{friedman2008sparse,
  title={Sparse inverse covariance estimation with the graphical lasso},
  author={Friedman, Jerome and Hastie, Trevor and Tibshirani, Robert},
  journal={Biostatistics},
  volume={9},
  number={3},
  pages={432--441},
  year={2008},
  publisher={Oxford University Press}
}

@article{yuan2007model,
  title={Model selection and estimation in the Gaussian graphical model},
  author={Yuan, Ming and Lin, Yi},
  journal={Biometrika},
  volume={94},
  number={1},
  pages={19--35},
  year={2007},
  publisher={Oxford University Press}
}

@article{cheng2014sparse,
  title={A sparse Ising model with covariates},
  author={Cheng, Jie and Levina, Elizaveta and Wang, Pei and Zhu, Ji},
  journal={Biometrics},
  volume={70},
  number={4},
  pages={943--953},
  year={2014},
  publisher={Wiley Online Library}
}

@article{ravikumar2010high,
  title={High-dimensional Ising model selection using $\ell_1$-regularized logistic regression},
  author={Ravikumar, Pradeep and Wainwright, Martin J and Lafferty, John D},
  journal={The Annals of Statistics},
  volume={38},
  number={3},
  pages={1287--1319},
  year={2010},
  publisher={Institute of Mathematical Statistics}
}

@article{hofling2009estimation,
  title={Estimation of sparse binary pairwise markov networks using pseudo-likelihoods.},
  author={H{\"o}fling, Holger and Tibshirani, Robert},
  journal={Journal of Machine Learning Research},
  volume={10},
  number={4},
  year={2009}
}

@article{wang2011learning,
  title={Learning oncogenic pathways from binary genomic instability data},
  author={Wang, Pei and Chao, Dennis L and Hsu, Li},
  journal={Biometrics},
  volume={67},
  number={1},
  pages={164--173},
  year={2011},
  publisher={Wiley Online Library}
}

@article{ravikumar2011high,
  title={High-dimensional covariance estimation by minimizing $\ell_1$-penalized log-determinant divergence},
  author={Ravikumar, Pradeep and Wainwright, Martin J and Raskutti, Garvesh and Yu, Bin},
  journal={Electronic Journal of Statistics},
  volume={5},
  pages={935--980},
  year={2011},
  publisher={The Institute of Mathematical Statistics and the Bernoulli Society}
}

@article{fierst2015modeling,
  title={Modeling the evolution of complex genetic systems: The gene network family tree},
  author={Fierst, Janna L and Phillips, Patrick C},
  journal={Journal of Experimental Zoology Part B: Molecular and Developmental Evolution},
  volume={324},
  number={1},
  pages={1--12},
  year={2015},
  publisher={Wiley Online Library}
}

@article{geman1993stochastic,
  title={Stochastic relaxation, Gibbs distributions and the Bayesian restoration of images},
  author={Geman, Stuart and Geman, Donald},
  journal={Journal of Applied Statistics},
  volume={20},
  number={5-6},
  pages={25--62},
  year={1993},
  publisher={Taylor \& Francis}
}

@article{marsman2018introduction,
  title={An introduction to network psychometrics: Relating Ising network models to item response theory models},
  author={Marsman, M and Borsboom, D and Kruis, J and Epskamp, S and Van Bork, R and Waldorp, LJ and Maas, HLJ van der and Maris, G},
  journal={Multivariate Behavioral Research},
  volume={53},
  number={1},
  pages={15--35},
  year={2018},
  publisher={Taylor \& Francis}
}

@article{ahsan1998elasticity,
  title={Elasticity theory of the B-DNA to S-DNA transition},
  author={Ahsan, Amir and Rudnick, Joseph and Bruinsma, Robijn},
  journal={Biophysical Journal},
  volume={74},
  number={1},
  pages={132--137},
  year={1998},
  publisher={Elsevier}
}

@article{dawid1979conditional,
  title={Conditional independence in statistical theory},
  author={Dawid, A Philip},
  journal={Journal of the Royal Statistical Society: Series B (Methodological)},
  volume={41},
  number={1},
  pages={1--15},
  year={1979},
  publisher={Wiley Online Library}
}

@article{delano2002pymol,
  title={The PyMOL Molecular Graphics System. De-Lano Scientific, San Carlos, CA, USA},
  author={DeLano, Warren L},
  journal={http://www. pymol. org},
  year={2002}
}
\end{document}